\documentclass[runningheads]{llncs}

\usepackage[T1]{fontenc}
\usepackage{graphicx}
\usepackage{amsmath,amssymb}
\usepackage{booktabs}
\usepackage{xcolor}
\usepackage{subcaption}
\usepackage[hidelinks]{hyperref}
\usepackage{cleveref}
\usepackage{placeins}
\usepackage{eso-pic}

\graphicspath{{figures/}}

\title{Collective Sensing\texorpdfstring{\\}{ }as Emergent Bayesian Inference}

\titlerunning{Collective Sensing as Emergent Bayesian Inference}
\author{Conor Heins\inst{1,2}\orcidID{0000-0002-5884-7728}}
\authorrunning{C. Heins}
\institute{Independent Researcher \and
Department of Collective Behaviour, Max Planck Institute of Animal Behavior, Konstanz, Germany
\email{conor.heins@gmail.com}}

\begin{document}
\maketitle

% Preprint-only acceptance notice. The original manuscript layout is retained.
\AddToShipoutPictureFG*{%
  \AtPageLowerLeft{%
    \put(0,42){\makebox[\paperwidth][c]{%
      \parbox{\textwidth}{\centering\footnotesize
        Accepted for publication in the proceedings of the\\
        7th International Workshop on Active Inference (IWAI 2026).}%
    }}%
  }%
}

\begin{abstract}
Collective behavior in animal groups is routinely described in cognitive language: fish schools sense gradients, swarms compute, colonies decide. These formulations often leave open what the collective is inferring, what computational object is represented, or how that representation arises from microscopic interactions. Here we develop that description quantitatively in a simulation based on the canonical model of collective sensing in fish schools (Berdahl et al., 2013), in which collective gradient sensing emerges from individuals that do not estimate gradients themselves: agents only modulate their speed in response to local light intensity and respond to simple social forces from neighbors. We show that these local rules admit a mesoscopic Bayesian reinterpretation at the level of the school, and hypothesize that the collective dynamics realize approximate inference on a generative model of how darkness gradients drive inter-individual speed differences. The resulting school-level inference process yields a posterior distribution over the local darkness gradient, whose statistics vary with environmental structure and relate systematically to collective motion and sensing performance. The school can therefore be described, in a Bayesian-mechanical sense, as a distributed inferential system without requiring individual agents to perform probabilistic belief updates themselves. This yields a quantitative, falsifiable account of collective sensing as emergent inference, and a concrete bridge from collective behavior to Bayesian mechanics.

\end{abstract}

% =====================================================================
\section{Introduction}\label{sec:intro}

The free-energy principle asks under what conditions random dynamical systems can be described as acting as Bayesian, evidence-maximizing systems~\cite{friston2010,friston2019,ramstead2018}. Two
complementary strands have grown from it. \emph{Bayesian mechanics} is
descriptive: it interprets the dynamics of certain classes of systems
as self-evidencing -- behaving as if they maximized the evidence for an
internal generative model of their sensory exchange with the
world~\cite{hohwy2016,friston2019} -- and remains an area of
theoretical work on the physics of self-organizing
systems~\cite{dacosta2021,ramstead2023}. Active inference is the
applied process theory built on the same objective: it is used to
construct agents -- whether as models of natural systems or to control artificial ones -- that act to minimize variational and expected free
energy~\cite{dacosta2020,parr2022}.

The free energy principle has now seen various applications to the study of collective behavior and multi-agent systems. The most common manifestation of this research direction usually takes the form of analyzing the collective behavior of multiple, interacting active inference agents. This framework has been used to investigate multi-agent cooperation~\cite{pitliya2025theory}, communication~\cite{friston2024federated,albarracin2022epistemic}, collective foraging by ant colonies~\cite{friedman2021}, morphogenesis and pattern formation~\cite{kuchling2020} and collective motion~\cite{heins2024,maisto2025}, among others.
A less common but complementary approach instead analyzes the collective itself as an emergent active inference agent. Waade et al.~\cite{waade2025} empirically investigate a system in which a population of low-level active inference agents can be mapped to a single group-level active inference agent; Heins et al.~\cite{heins2022spinglass} derive such a mapping analytically in the Ising model, showing that a network of coupled spins -- each cast as an active inference agent -- collectively performs inference under a group-level Boltzmann machine generative model.

The empirical study of
collective behavior, meanwhile, routinely attributes agentic computations to
biological collectives -- they ``sense'', ``decide'', ``act'', and
``estimate''~\cite{couzin2009,couzin2002}. But that attribution is usually phenomenological: it is
established by quantifying properties of the group behavior -- such as observing that collective performance
on some task varies with properties of the collective system like its size or topology -- rather than by explicitly
identifying the computational mechanism implemented by the group. Notable exceptions cast swarms as collective probability estimators~\cite{hunt2020,krafft2021}, but these typically endow each individual with a local probabilistic model that is a reduced, local version of the group-level model. Even recent attempts to read a flock as an emergent agent characterize it through information-theoretic
signatures of joint computation rather than an identified group-level
generative model~\cite{maisto2025}.
Specifying the inference \emph{formally} would allow us to explicitly map collective dynamics to the  computations underlying intelligent group behavior. This paper argues that Bayesian mechanics and active inference provide a useful framework for developing this formal connection.

In particular, this paper provides a worked example of deriving a Bayesian inferential interpretation of collective sensing in animal groups. Such groups routinely achieve sensing and
decisions beyond the reach of any individual member, throwing the gap
between individual rules and group-level computation into sharp relief~\cite{couzin2009,einolghozati2012collective,nagy2018synchronization,pedraja2024collective,franks2002information}.
We study the gradient-tracking schooling model of Berdahl et
al.~\cite{berdahl2013}, which captures how schools of golden shiners
(\emph{Notemigonus crysoleucas}) -- small fish that prefer dim water but
cannot individually estimate the ambient darkness gradient -- collectively
locate and track moving regions of shade. The model represents these fish as
agents that modulate their speed in response to the local light intensity and follow
simple social attraction and alignment rules, yet the school as a whole
tracks a moving dark region (the same speed-up/slow-down principle later
engineered into distributed source-seeking
controllers~\cite{alabri2018susd}). We ask the inverse question: which generative model and associated inference and belief-guided action describe the collective's dynamics? We construct a school-level generative model that yields a posterior belief about the local darkness gradient, and hypothesize that inference on this model is realized through the school's collective dynamics. We then relate the posterior to the school's motion and sensing behavior.

We examine three aspects of this account. First, the school-level posterior recovers gradient directions, with uncertainty that depends on sampling geometry and residual correlations. Second, perturbations of the directional evidence and changes in the light field test its environmental coupling. Third, a heading model combining the inferred gradient with persistence predicts collective motion, and a fitted relation links these headings to the group-size dependence of sensing performance~\cite{berdahl2013}.

Throughout this work, we adopt a particular parameterization of the proposed generative model and assess its gradient estimates and behavioral predictions. The account is falsifiable through its likelihood, uncertainty estimates, and predicted relation to motion. We also derive conditions under which the inferred gradient's contribution to the observed group heading takes the form of a one-step expected-free-energy action (Supplementary \Cref{sec:supp:efe}). 

\begin{figure}[tbp]
\centering
\includegraphics[width=0.82\textwidth]{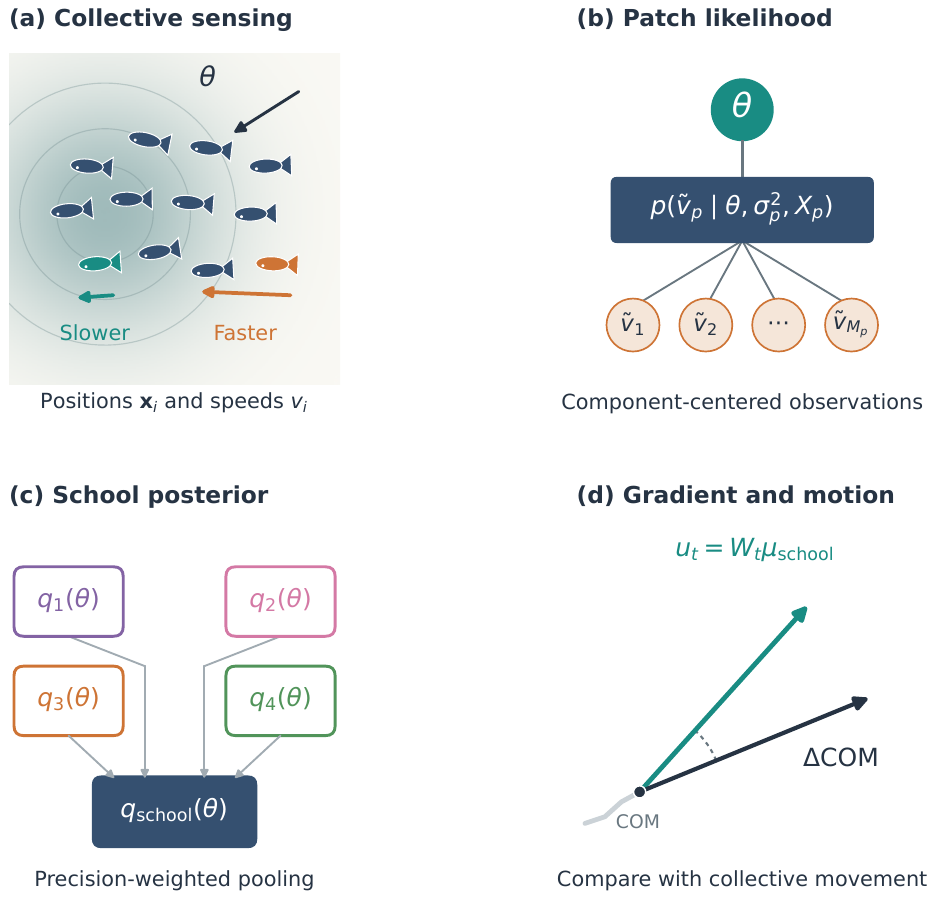}
\caption{From school observations to gradient and heading estimates. \textbf{(a)}~A
school moving through a 2D darkness field. Example velocity arrows show the light-dependent speed response; $\theta$ denotes the local darkness gradient. \textbf{(b)}~The factor-graph patch
model: centered speeds~$\tilde v_i$ provide evidence about~$\theta$ under a Gaussian working likelihood with state-dependent covariance (\Cref{eq:school-coupled-likelihood}). \textbf{(c)}~Patch posteriors combine by federated inference into the
school posterior, a precision-weighted natural-parameter sum.
\textbf{(d)}~Alignment between the school's actual frame-to-frame
center-of-mass displacement~$\Delta\mathrm{COM}$ and the
posterior-gradient drive $u_t = W_t\,\mu_\mathrm{school}$
(\Cref{eq:action}). The fitted heading model also includes directional persistence.}
\label{fig:pipeline}
\end{figure}

% =====================================================================
\section{Schooling model, emergent generative model, inference, and action}\label{sec:methods}

\subsection{Per-agent dynamics and the light field}\label{sec:methods:dynamics}

We use a simulation based on Berdahl et al.~\cite{berdahl2013} and zone-based social interactions~\cite{couzin2002}. Agent~$i$ has position~$\mathbf{x}_i$, unit heading~$\mathbf{d}_i$, and speed~$v_i$, with
$\mathbf{x}_i(t{+}1)=\mathbf{x}_i(t)+v_i\mathbf{d}_i\Delta t$.
Speed increases with the local light value: $v_i=v_\mathrm{min}+(v_\mathrm{max}-v_\mathrm{min})g_t(L(\mathbf{x}_i,t))$, where 
$g_t$ normalizes $L$ to $[0,1]$ using the field's minimum and maximum at time $t$. Heading combines alignment, attraction, repulsion, and noise, with repulsion taking priority at short range~\cite{reynolds1987flocks,couzin2002}. Schematically, outside the repulsion zone,
\begin{equation}
\mathbf{d}_i(t{+}1)\propto w_\mathrm{ori}\mathbf{r}^\mathrm{ori}_i+w_\mathrm{att}\mathbf{r}^\mathrm{att}_i+\boldsymbol{\xi}_i,
\label{eq:heading}
\end{equation}
where $\mathbf{r}_i$ denotes a social direction and $\boldsymbol{\xi}_i$ heading noise. Supplementary \Cref{sec:supp:simulation} gives the full update rules and simulation parameters.

We write darkness as $D$ and its complementary light coordinate as $L=1-D$. The field combines a moving dark feature and a correlated background,
\begin{equation}
D(\mathbf{x},t)=\max\!\left(f_\mathrm{min},f_\mathrm{max}e^{-\|\mathbf{x}-\mathbf{p}(t)\|/\ell_\mathrm{field}}\right)+\alpha b(\mathbf{x},t).
\label{eq:light}
\end{equation}
Here $\mathbf{p}(t)$ is the dark-feature center, $\ell_\mathrm{field}$ its exponential decay length, and $b$ a Gaussian background with spectral parameter $k_0$ and temporal relaxation parameter $\nu$. Positive $f_\mathrm{max}$ raises darkness, slowing agents through the normalized speed response.

\subsection{The generative model}\label{sec:methods:model}

In this section we propose an `emergent' generative model for the school; under this proposal, approximate inference and action lead to behavior consistent with the explicit equations of motion given in Section~\ref{sec:methods:dynamics}.

The model connects perception and action through two latent variables: the darkness gradient $\theta_t=\nabla D=-\nabla L\in\mathbb{R}^2$ and a collective heading state $h_t\in\mathbb{R}^2$. The direction of $h_t$ represents collective action, with its magnitude scaled into gradient units. The observations are speeds $v_{p,t}$ in spatial patch~$p$ at time~$t$:
\begin{equation}
p(v,\theta_{1:T},h_{1:T})=\prod_t p(h_t\mid h_{t-1})p(\theta_t\mid h_t)\textstyle\prod_p p(v_{p,t}\mid\theta_t).
\label{eq:genmodel}
\end{equation}
Heading persists around the previous heading direction (normalized to unit magnitude $\hat{h}_t$, while the second factor couples the heading to the light gradient:
\begin{equation}
p(h_t\mid h_{t-1})=\mathcal{N}(\rho\widehat h_{t-1},\Sigma_h),\qquad
p(\theta_t\mid h_t)=\mathcal{N}(h_t,\Sigma_c I),\qquad \widehat u=u/\|u\|.
\label{eq:heading-transition}
\end{equation}
Here $I$ is the $2\times2$ identity, $\rho$ the persistence strength, $\Sigma_h$ the heading covariance, and $\Sigma_c>0$ the coupling variance. Although $p(\theta_t\mid h_t)$ is written in the generative direction, during inference the evidence accumulation for the gradient variable comes to influence the heading. Its variance controls the coupling strength. We hypothesize that social turning and speed differences realize this inference process (see Supplementary \Cref{sec:supp:assumptions} for the assumptions underlying this).

\paragraph{Gradient likelihood.}
Speed modulation makes differences between spatially separated agents informative about the light gradient. Social coupling shapes their sampling positions and turns these speed differences into collective movement, motivating the likelihood below.

Within a defined spatial patch of light, we can linearize the light value at a given position $\mathbf{x}_i$ with
$L(\mathbf{x}_i)\approx\bar L-\theta^\top(\mathbf{x}_i-\bar{\mathbf{x}})$,
where $\theta=-\nabla L$ and the bars denote averages over a local group of agents. We define patches as overlapping circular regions that cover the school's extent and are aligned with its heading. Each has radius $R=\max(10,0.4S)$, where $S$ is the farthest agent's distance from the school center of mass. Their centers lie at the center of mass, $1.5R$ ahead along the mean heading, and $1.2R$ to either side, shifted inward near arena boundaries. Within each patch, we define groups of agents by connected components of a proximity graph, where an edge between any two agents is defined by whether those agents are separated by at most $\frac{r_o}{2}$. Speed differences are then treated separately within these patch-specific connected components.

Taking speed differences within a component removes its common mean response and gives, to leading order,
\begin{equation}
v_i - v_j \;\approx\; \kappa_v\,\theta^\top(\mathbf{x}_i - \mathbf{x}_j)
 + (\varepsilon_i - \varepsilon_j).
\label{eq:pairwise}
\end{equation}
Here $\kappa_v=\partial v/\partial D<0$ is the local speed sensitivity, assumed known, and $\varepsilon_i$ a residual capturing departures from the linear relation. The pairwise differences contain the same centered information as regressing each agent's centered speed
$\tilde v_i = v_i - \bar v$ on its centered position
$x_i = \mathbf{x}_i - \bar{\mathbf{x}}$, with both means taken over the component. Centering removes the unknown
component-mean speed; stacking component-centered observations gives a linear-Gaussian patch likelihood,
\begin{equation}
\tilde v \mid \theta, \sigma^2, X \;\sim\; \mathcal{N}\!\big(\kappa_v\,X\theta,\;\sigma^2 K(X)\big).
\label{eq:school-coupled-likelihood}
\end{equation}
The rows of $X$ are centered within each component, and $\sigma^2$ is the residual variance scale. Social coupling shapes positions and trajectories through heading, so agents sample correlated departures from the local linear field approximation. We model these residual speed correlations with a Gaussian working covariance (Supplementary \Cref{app:centering}). Within component~$k$ of patch~$p$,
\begin{equation}
K_{p,k;ij} = \exp\!\left(-\frac{\|x_i - x_j\|^2}{2\ell_{p,k}^2}\right),
\qquad
\ell_{p,k} = c\,\sqrt{\tfrac{1}{N_{p,k}}\textstyle\sum_{i\in\mathcal C_{p,k}} \|x_i\|^2}.
\label{eq:rbf-kernel_and_ell-eff}
\end{equation}

Here $\mathcal C_{p,k}$ contains the $N_{p,k}$ agents in component $k$ and dimensionless $c$ scales its RMS spread. With $C_p$ retained components, $K_p=\operatorname{blockdiag}(K_{p,1},\ldots,K_{p,C_p})$ has blocks of size $N_{p,k}\times N_{p,k}$. Write $K=K_p+10^{-6}I$ to include diagonal jitter, and define
$G_p=\kappa_v^2X^\top K^{-1}X$, $g_p=\kappa_vX^\top K^{-1}\tilde v$, and $\widetilde G_p=G_p+0.02I$ to regularize unmeasured directions.
Empirical Bayes (EB) estimates residual precision from these data using inverse-gamma variance hyperparameters $a_0,b_0$:
\begin{equation}
\widehat\sigma_p^{-2}=\frac{a_0+M_p/2}{b_0+\tfrac12(\tilde v^\top K^{-1}\tilde v-g_p^\top\widetilde G_p^{-1}g_p)}.
\label{eq:eb-noise}
\end{equation}
$M_p=\sum_{k=1}^{C_p}N_{p,k}$ counts retained observations. The regularized Gaussian message, combined with a weak isotropic prior $\Lambda_0$, gives the static patch posterior,
\begin{equation}
\Lambda_p=\Lambda_0+\widehat\sigma_p^{-2}\widetilde G_p,\qquad
\mu_p=\Lambda_p^{-1}\widehat\sigma_p^{-2}g_p.
\label{eq:patch-posterior}
\end{equation}
Larger $b_0$ lowers the estimated precision. Supplementary \Cref{sec:supp:assumptions,app:iid} describe the implemented temporal filtering and EB approximation.

\subsection{Patch-federated inference}\label{sec:methods:inference}

We first estimate the gradient from speed evidence, neglecting feedback from the heading prior in \eqref{eq:genmodel}. Conditional on plug-in precisions, the equation for the shared posterior over $\theta_t$ is a federated inference step that sum independent patch messages, yielding a shared gradient estimate:
\begin{equation}
\begin{split}
\Lambda_\mathrm{school}&=\Lambda_0+\sum_{p\in\mathcal A}\widehat\sigma_p^{-2}\widetilde G_p,\\
\mu_\mathrm{school}&=\Lambda_\mathrm{school}^{-1}\sum_{p\in\mathcal A}\widehat\sigma_p^{-2}g_p,\qquad
\Sigma_\mathrm{school}=\Lambda_\mathrm{school}^{-1}.
\end{split}
\label{eq:federation}
\end{equation}
$\mathcal A$ contains patches with at least one retrained connected component: retained components are those with two or more agents and whose maximum distance from their mean position is at least one arena unit. The eigenvalues of $G_p$ determine which gradient directions are constrained. Pooling assumes a common gradient and distinct evidence; overlapping windows, temporal correlations, and spatial variation naturally threaten this approximation.

The heading estimate balances its previous direction against the gradient message:
\begin{equation}
\begin{split}
m_t^h&=(\Sigma_h^{-1}+W_t)^{-1}(\Sigma_h^{-1}\rho\widehat h_{t-1}+W_t\mu_\mathrm{school}),\\
W_t&=(\Sigma_c I+\Sigma_\mathrm{school})^{-1}.
\end{split}
\label{eq:heading-update}
\end{equation}
$W_t$ weights directions by the precision of the gradient message, capped by $\Sigma_c^{-1}$. We use this conditional Gaussian mean to predict the school's next heading, taking the direction of its previous observed displacement as $\widehat h_{t-1}$.

\subsection{Acting on the posterior}\label{sec:methods:action}

The gradient contribution to the heading update is
\begin{equation}
u_t=W_t\mu_\mathrm{school}=(\Sigma_c I+\Sigma_\mathrm{school})^{-1}\mu_\mathrm{school}.
\label{eq:action}
\end{equation}
This posterior-gradient drive strengthens well-constrained components and attenuates uncertain ones. It is parallel to the posterior mean when the total covariance is isotropic, and approaches that direction when $\Sigma_c I$ dominates. The full predicted heading $\widehat m_t^h$ also contains persistence and generally differs from $\widehat u_t$.

This form also arises in a one-step expected-free-energy (EFE) calculation with a preference for low light. In a locally linear, quasi-static field, the objective relates a proposed displacement to the expected light level and its uncertainty. Supplementary \Cref{sec:supp:efe} derives the conditions on the covariance and constrained optimum that yield the precision-weighted direction in \eqref{eq:action}. This supplies an action hypothesis linking the school posterior to movement toward darkness.

\subsection{Simulation experiments}\label{sec:methods:experiments}

We simulate seven group sizes, doubling from $N=4$ to $N=256$, with 64 seeds at each size. Training seeds determine the heading parameters and the mapping to sensing performance; a complementary split selects the posterior parameters, giving $\phi^\star=(c,b_0,\rho,\Sigma_c)$ (Supplementary \Cref{sec:results:inverse}). We compare posterior means and uncertainty with the known simulated gradients, and predicted headings with subsequent school motion. Supporting analyses include simulation-based calibration of simplified likelihoods~\cite{talts2018}, perturbations of the evidence supplied to the estimator, and sweeps over environmental structure. Supplementary \Cref{sec:supp:simulation} lists the likelihood variants and parameters used in each analysis.

% =====================================================================
\section{Results}\label{sec:results}

We test the school-level inference interpretation along three axes: gradient estimation and uncertainty, coupling to the environment, and behavioral relevance. We begin with the information supplied by a few agents, then examine larger schools and their motion and sensing performance.

\subsection{The primitive: speed differences identify gradient components}\label{sec:results:primitive}

\begin{figure}[tbp]
\centering
\includegraphics[width=\textwidth]{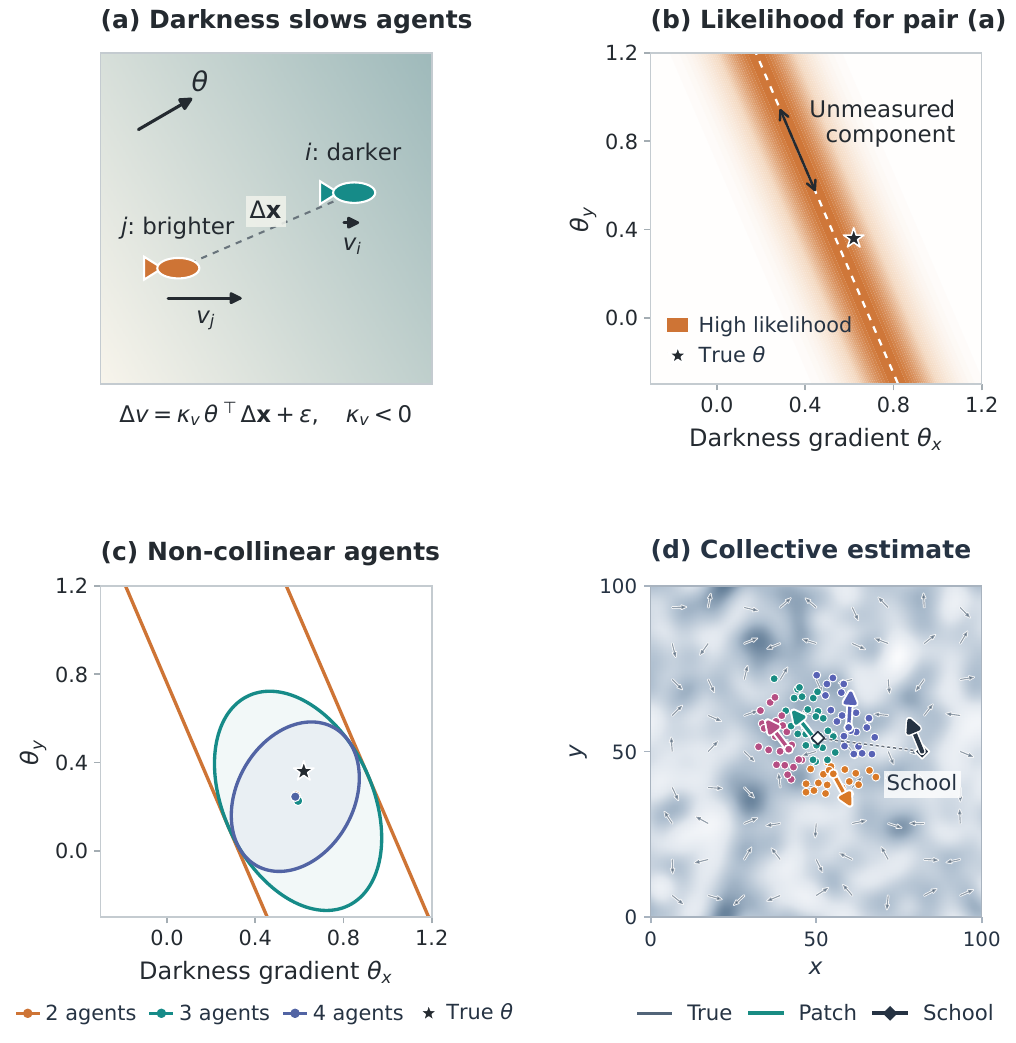}
\caption{From local speed differences to a collective gradient estimate.
\textbf{(a)}~In a linear darkness field, the agent in darker water moves more slowly:
$\Delta v=\kappa_v\theta^\top\Delta\mathbf{x}+\varepsilon$, with $\kappa_v<0$.
\textbf{(b)}~Likelihood for the same pair in (a), plotted in gradient space. Orange shading shows relative likelihood; the dashed ridge follows gradients with the same predicted speed difference. Its tangent is perpendicular to the pair's spatial separation.
\textbf{(c)}~Adding non-collinear agents yields compact 95\% posterior regions; dots are means and the star is the true gradient. The same weak prior closes the two-agent contour beyond the displayed range.
\textbf{(d)}~A structured-field school ($N=96$, frame~33; $c=0.35$, $b_0=0.5$), chosen for arrow separation. Darker shading denotes greater darkness. Dots identify agents by a patch assignment; windows can overlap. Thin arrows show true directions and colored arrows patch estimates. The black school arrow is offset for clarity, with a dashed leader to the diamond at the school centroid. Arrows indicate direction only. Mean cosine at patch centroids is $0.735$.}
\label{fig:twofish}
\end{figure}

The basic computation is visible in the pair in \Cref{fig:twofish}a. Their speed difference measures the gradient along their separation (\Cref{eq:pairwise}). Panel~b plots this pair's likelihood over possible gradients $(\theta_x,\theta_y)$: values along the orange ridge give the same predicted speed difference, so its direction is unmeasured. Adding non-collinear agents supplies another projection, contracting the posterior in panel~c. Panel~d applies the same construction to a simulated school in a spatially varying field.

\subsection{The selected generative model: calibration and the group-size law}\label{sec:results:calibration}\label{sec:results:berdahl}

\begin{figure}[tbp]
\centering
\includegraphics[width=0.95\textwidth]{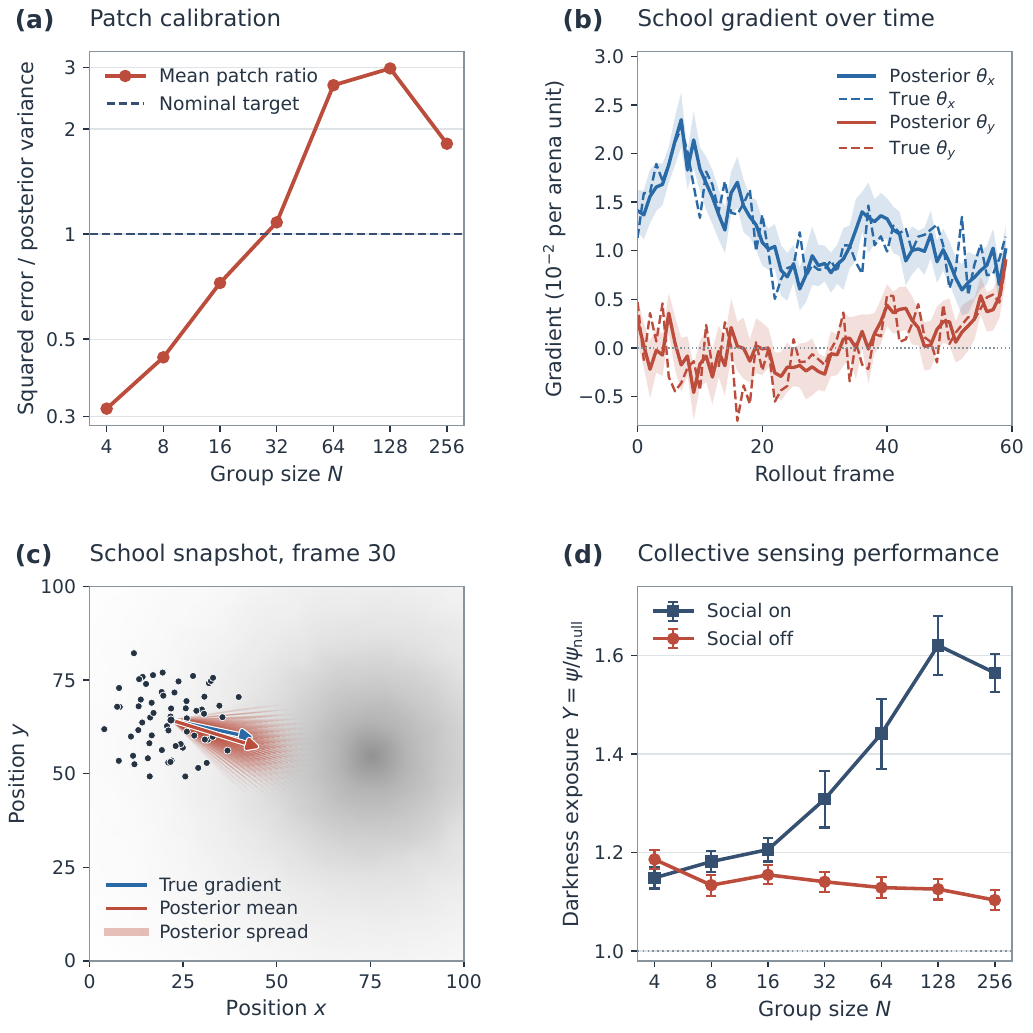}
\caption{Gradient inference and sensing performance at the selected parameters ($c=0.35$, $b_0=4.0$).
\textbf{(a)}~Mean active-patch error-to-variance ratio vs.\ group size~$N$ (16 seeds per size). Both reference and posterior gradients use arena units. The dashed line marks the nominal target of one.
\textbf{(b)}~A single $N=64$ rollout: the
horizontal axis is rollout frame; solid curves show posterior-mean $\theta_x,\theta_y$, dashed curves the truth, and shading marginal $\pm1\sigma$ uncertainty. \textbf{(c)}~A frame snapshot: agent positions, the
true darkness gradient, and the posterior-mean gradient with its
uncertainty envelope (faded arrows, opacity $\propto$ posterior
density). \textbf{(d)}~The darkness-exposure ratio $Y$
vs.~$N$ for canonical coupling and the social-off ablation (means and standard errors over 64 seeds). The ablation sets alignment and attraction weights to zero while retaining light-dependent speed and repulsion; implementation details are in Supplementary \Cref{sec:supp:simulation}.}
\label{fig:chosen}
\label{fig:berdahl}
\end{figure}

We next estimate gradients in schools of different sizes using the covariance kernel in \eqref{eq:rbf-kernel_and_ell-eff}. The selected parameters are $c=0.35$ and $b_0=4.0$, with the fitting procedure described in Supplementary \Cref{sec:results:inverse}. We measure directional accuracy by $\cos(\mu,\theta)=\frac{\mu^\top\theta}{\|\mu\|\|\theta\|}$, evaluating the true gradient at each patch center and at the school's center of mass (COM). Mean patch cosine ranges from $0.76$ to $0.96$ across group sizes. School cosine is $0.97$ at $N=64$ and $0.54$ at $N=256$, indicating that agreement with the local field can coexist with substantial variation in direction across a large school. A single rollout illustrates the inferred gradient over time (\Cref{fig:chosen}b), and a snapshot shows the estimate and its uncertainty at the school COM (\Cref{fig:chosen}c).

To assess uncertainty, we compare squared estimation error with posterior variance through the ratio $\frac{|\theta_{p,\mathrm{true}}-\mu_p|^2}{\mathrm{Tr}(\Sigma_p)}$. This ratio has expectation one when the estimation error follows the stated Gaussian posterior. Values above one indicate overconfidence, and values below one indicate a posterior wider than the observed errors. \Cref{fig:chosen}a shows this diagnostic averaged over active patch--frame observations at each group size, with reference and inferred gradients expressed in arena units. The mean ratio rises above one at $N=32$ and reaches $2.98$ at $N=128$, indicating that the posterior becomes overconfident as the school grows. The model comparison motivated the correlated likelihood: pairwise differences share observations, and residual correlations persist after centering (Supplementary \Cref{sec:supp:calibration}).

Permuting speeds within local patches reduces directional accuracy to near zero; Supplementary \Cref{sec:supp:doseresponse} reports these evidence controls and an analysis of uncertainty under overlapping and exclusive patch membership. Environmental sweeps examine the effects of background noise and spatial scale (Supplementary \Cref{sec:results:fieldnoise,sec:results:boundary,sec:results:falsification}).

The sensing-performance curves are measured from the simulator, independently of posterior inference. Following Berdahl et
al.~\cite{berdahl2013}, we measure collective sensing through the
darkness-exposure ratio $Y=\frac{\psi}{\psi_\mathrm{null}}$ where $\psi=\langle
1-L_\mathrm{norm}\rangle_{i,t}$ is the school's time- and agent-averaged darkness (higher is
darker), with $L_\mathrm{norm}$ the complement of normalized darkness. $\psi_\mathrm{null}$ uses an independently seeded reference field; $Y>1$ means greater darkness exposure than this control,
and the central result is that $Y$ rises with~$N$ for coupled agents. Our pipeline
shows $Y$ rising with~$N$ at the canonical social weights and a flatter curve
when social coupling is ablated (\Cref{fig:berdahl}d). This `social off' configuration sets alignment and attraction weights to zero while retaining light-dependent speed and repulsion. The non-monotonic
dependence on the alignment/attraction balance
$r=\frac{w_\mathrm{ori}}{w_\mathrm{ori}+w_\mathrm{att}}$ is reproduced too
(\Cref{sec:supp:berdahl}).

\subsection{Action: collective motion and the inferred posterior}\label{sec:results:efe}\label{sec:results:heading}

\begin{figure}[tbp]
\centering
\includegraphics[width=0.88\textwidth]{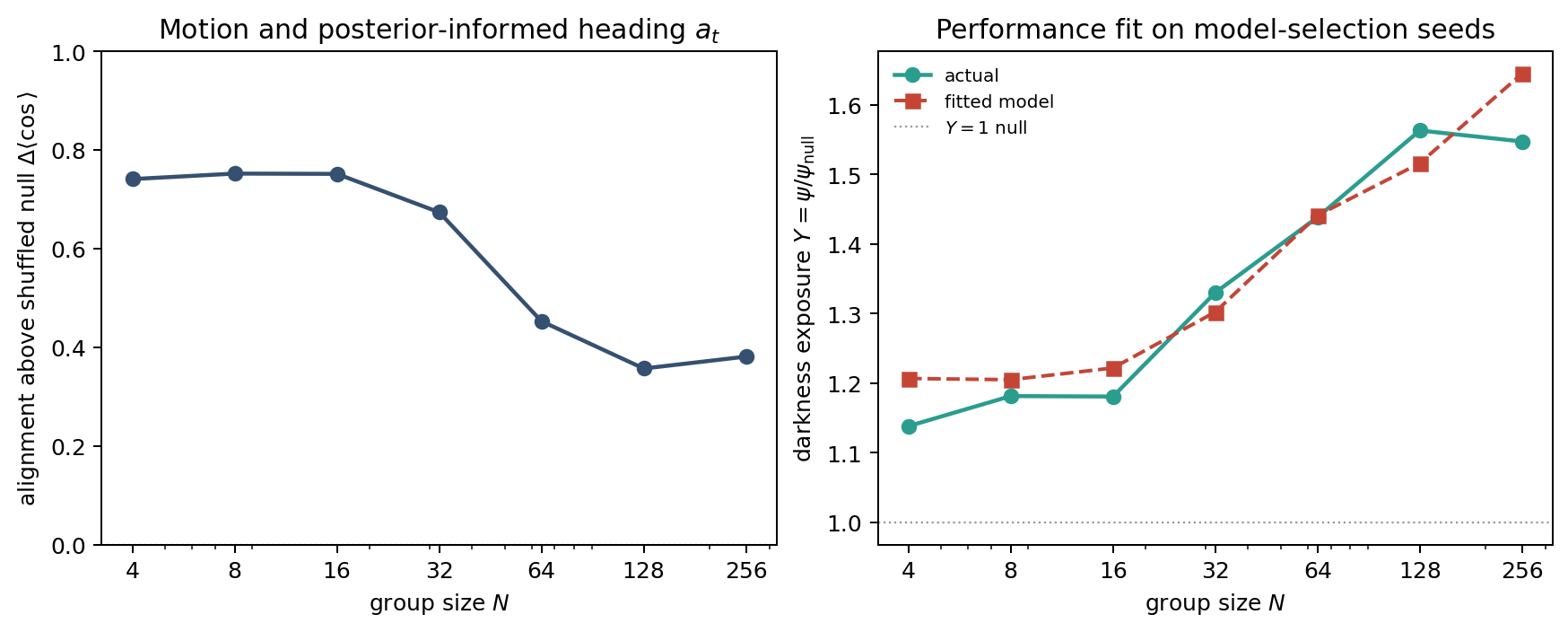}
\caption{Collective motion and sensing performance in the fitted model. \textbf{(a)}~Alignment between displacement and the predicted heading, including persistence, relative to the shuffled-action baseline. \textbf{(b)}~The fitted relation $\widehat Y(N)$ alongside observed $Y(N)$ and the $Y=1$ reference. Both panels show the evaluation split used to select the posterior parameters.}

\label{fig:action}
\end{figure}

We use the gradient posterior and the previous direction of motion to predict the school's next heading, $a_t=\widehat m_t^h$ (\Cref{eq:heading-update}). \Cref{fig:action}a compares this heading with the subsequent COM displacement, reporting mean cosine relative to a within-condition shuffled-action baseline. Alignment is strongest for small schools and declines with $N$. Collective movement combines responses to the field with persistence and social interactions, so agreement with the instantaneous gradient depends on the spatial structure of the field and the school's recent trajectory.

We then ask how the model's heading estimates relate to sensing performance. Panel~b shows the positive affine regression $\widehat Y=A+B\langle\cos(a_t,\theta_\mathrm{COM})\rangle$, $B\geq0$, relating their mean alignment with the true simulated COM gradient to the observed darkness-exposure ratio. We fit the coefficients on training seeds and apply them to the complementary model-selection split, obtaining $R^2\approx0.90$. The relation thus links the model's directional estimates to the observed group-size dependence of sensing. Supplementary \Cref{sec:results:inverse} gives the regression and selection procedure.

% =====================================================================
\FloatBarrier
\section{Discussion}\label{sec:discussion}
We developed and tested an emergent Bayesian interpretation of collective sensing in the gradient-tracking model introduced by Berdahl et al.~\cite{berdahl2013}. Under the proposed model, the school encodes a federated posterior over the darkness gradient whose directions agree with the local field and whose statistics vary with environmental structure. The posterior also relates systematically to collective motion and sensing performance. This provides a quantitatively grounded description of collective sensing in terms of distributed Bayesian inference.

In this interpretation, the posterior is a property of the school as a coarse-grained system and belief-guided action a property of its \emph{macroscopic} dynamics. The inference is distributed across the members' speed differences and interactions. We therefore treat ``is the collective doing inference?'' as an \emph{inverse} problem: we seek a mechanistically plausible generative model under which the school can be understood as both encoding an environmentally sensitive posterior and acting on it. The resulting hypothesis gives falsifiable predictions for how changes in local sensing, geometry, and social coupling affect inference and movement. With the model and parameters held fixed, systematic errors in gradient direction, uncertainty that fails to track estimation error, or failures to predict the response of subsequent motion to these changes would count against the account within the regime being tested.

\paragraph{Limitations and future work.} The current tests compare an estimated posterior with simulated gradients and behavior. A direct dynamical test would identify collective physical quantities corresponding to the belief parameters, including the state that carries temporal information, and examine whether their evolution follows the proposed approximate-inference updates across inputs and perturbations (Supplementary \Cref{sec:supp:assumptions}). The likelihood and patch partition are modeling choices: local field curvature, overlapping observations and temporal correlations affect the inferred uncertainty, and school-level pooling remains overconfident in the supplementary analysis. Partial pooling and line-of-sight partitions~\cite{davidson2021collective} are natural extensions. Fresh simulations, with the selected likelihood and fitted parameters held fixed, would test predictive performance independently of the data used for model selection. Tests in other collective systems would help determine how far this account extends beyond the speed--light mechanism studied here. The EFE calculation supplies a further hypothesis for the relation between belief updates and collective action.

\clearpage
% Flush any pending main-text figures before the references so none
% float past the bibliography into the appendix.
\FloatBarrier
% =====================================================================
\bibliographystyle{splncs04}
\bibliography{refs}

\begin{thebibliography}{10}
\providecommand{\url}[1]{\texttt{#1}}
\providecommand{\urlprefix}{URL }
\providecommand{\doi}[1]{https://doi.org/#1}

\bibitem{alabri2018susd}
Al-Abri, S., Maxon, S., Zhang, F.: Integrating a {PCA} learning algorithm with
  the {SUSD} strategy for a collective source-seeking behavior. In: IEEE/RSJ
  International Conference on Intelligent Robots and Systems (IROS) (2018)

\bibitem{albarracin2022epistemic}
Albarracin, M., Demekas, D., Ramstead, M.J., Heins, C.: Epistemic communities
  under active inference. Entropy  \textbf{24}(4), ~476 (2022)

\bibitem{berdahl2013}
Berdahl, A., Torney, C.J., Ioannou, C.C., Faria, J.J., Couzin, I.D.: Emergent
  sensing of complex environments by mobile animal groups. Science
  \textbf{339}(6119),  574--576 (2013). \doi{10.1126/science.1225883}

\bibitem{couzin2009}
Couzin, I.D.: Collective cognition in animal groups. Trends in Cognitive
  Sciences  \textbf{13}(1),  36--43 (2009). \doi{10.1016/j.tics.2008.10.002}

\bibitem{couzin2002}
Couzin, I.D., Krause, J., James, R., Ruxton, G.D., Franks, N.R.: Collective
  memory and spatial sorting in animal groups. Journal of Theoretical Biology
  \textbf{218}(1),  1--11 (2002). \doi{10.1006/jtbi.2002.3065}

\bibitem{dacosta2020}
Da~Costa, L., Parr, T., Sajid, N., Veselic, S., Neacsu, V., Friston, K.: Active
  inference on discrete state-spaces: A synthesis. Journal of Mathematical
  Psychology  \textbf{99},  102447 (2020). \doi{10.1016/j.jmp.2020.102447}

\bibitem{dacosta2021}
Da~Costa, L., Friston, K., Heins, C., Pavliotis, G.A.: Bayesian mechanics for
  stationary processes. Proceedings of the Royal Society A  \textbf{477}(2256),
   20210518 (2021). \doi{10.1098/rspa.2021.0518}

\bibitem{davidson2021collective}
Davidson, J.D., Sosna, M.M., Twomey, C.R., Sridhar, V.H., Leblanc, S.P.,
  Couzin, I.D.: Collective detection based on visual information in animal
  groups. Journal of the Royal Society Interface  \textbf{18}(180) (2021)

\bibitem{einolghozati2012collective}
Einolghozati, A., Sardari, M., Fekri, F.: Collective sensing-capacity of
  bacteria populations. In: 2012 IEEE International Symposium on Information
  Theory Proceedings. pp. 2959--2963. IEEE (2012)

\bibitem{franks2002information}
Franks, N.R., Pratt, S.C., Mallon, E.B., Britton, N.F., Sumpter, D.J.:
  Information flow, opinion polling and collective intelligence in
  house--hunting social insects. Philosophical Transactions of the Royal
  Society of London. Series B: Biological Sciences  \textbf{357}(1427),
  1567--1583 (2002)

\bibitem{friedman2021}
Friedman, D.A., Tschantz, A., Ramstead, M.J.D., Friston, K., Constant, A.:
  Active inferants: An active inference framework for ant colony behavior.
  Frontiers in Behavioral Neuroscience  \textbf{15},  647732 (2021).
  \doi{10.3389/fnbeh.2021.647732}

\bibitem{friston2010}
Friston, K.: The free-energy principle: A unified brain theory? Nature Reviews
  Neuroscience  \textbf{11}(2),  127--138 (2010). \doi{10.1038/nrn2787}

\bibitem{friston2019}
Friston, K.: A free energy principle for a particular physics. arXiv preprint
  arXiv:1906.10184  (2019)

\bibitem{friston2024federated}
Friston, K.J., Parr, T., Heins, C., Constant, A., Friedman, D., Isomura, T.,
  Fields, C., Verbelen, T., Ramstead, M., Clippinger, J., et~al.: Federated
  inference and belief sharing. Neuroscience \& Biobehavioral Reviews
  \textbf{156},  105500 (2024)

\bibitem{heins2022spinglass}
Heins, C., Klein, B., Demekas, D., Aguilera, M., Buckley, C.L.: Spin glass
  systems as collective active inference. In: International Workshop on Active
  Inference (IWAI) (2022), arXiv:2207.06970

\bibitem{heins2024}
Heins, C., Millidge, B., Da~Costa, L., Mann, R.P., Friston, K.J., Couzin, I.D.:
  Collective behavior from surprise minimization. Proceedings of the National
  Academy of Sciences  \textbf{121}(17),  e2320239121 (2024).
  \doi{10.1073/pnas.2320239121}

\bibitem{hohwy2016}
Hohwy, J.: The self-evidencing brain. No\^{u}s  \textbf{50}(2),  259--285
  (2016). \doi{10.1111/nous.12062}

\bibitem{hunt2020}
Hunt, E.R., Franks, N.R., Baddeley, R.J.: The {Bayesian} superorganism:
  Collective probability estimation in swarm systems. In: Artificial Life
  Conference Proceedings. vol.~32, pp. 315--323 (2020)

\bibitem{krafft2021}
Krafft, P.M., Shmueli, E., Griffiths, T.L., Tenenbaum, J.B., Pentland, A.:
  {Bayesian} collective learning emerges from heuristic social learning.
  Cognition  \textbf{212},  104469 (2021).
  \doi{10.1016/j.cognition.2020.104469}

\bibitem{kuchling2020}
Kuchling, F., Friston, K., Georgiev, G., Levin, M.: Morphogenesis as {Bayesian}
  inference: A variational approach to pattern formation and control in complex
  biological systems. Physics of Life Reviews  \textbf{33},  88--108 (2020).
  \doi{10.1016/j.plrev.2019.06.001}

\bibitem{maisto2025}
Maisto, D., Nuzzi, D., Pezzulo, G.: What the flock knows that the birds do not:
  Exploring the emergence of joint agency in multi-agent active inference.
  arXiv preprint arXiv:2511.10835  (2025)

\bibitem{nagy2018synchronization}
Nagy, M., Couzin, I.D., Fiedler, W., Wikelski, M., Flack, A.: Synchronization,
  coordination and collective sensing during thermalling flight of freely
  migrating white storks. Philosophical Transactions of the Royal Society B:
  Biological Sciences  \textbf{373}(1746) (2018)

\bibitem{parr2022}
Parr, T., Pezzulo, G., Friston, K.J.: Active Inference: The Free Energy
  Principle in Mind, Brain, and Behavior. MIT Press (2022)

\bibitem{pedraja2024collective}
Pedraja, F., Sawtell, N.B.: Collective sensing in electric fish. Nature
  \textbf{628}(8006),  139--144 (2024)

\bibitem{pitliya2025theory}
Pitliya, R.J., {\c{C}}atal, O., Van~de Maele, T., Pezzato, C., Verbelen, T.:
  Theory of mind using active inference: A framework for multi-agent
  cooperation. In: International Workshop on Active Inference. pp. 385--400.
  Springer (2025)

\bibitem{ramstead2018}
Ramstead, M.J.D., Badcock, P.B., Friston, K.J.: Answering schr\"odinger's
  question: A free-energy formulation. Physics of Life Reviews  \textbf{24},
  1--16 (2018). \doi{10.1016/j.plrev.2017.09.001}

\bibitem{ramstead2023}
Ramstead, M.J.D., Sakthivadivel, D.A.R., Heins, C., Koudahl, M., Millidge, B.,
  Da~Costa, L., Klein, B., Friston, K.J.: On {B}ayesian mechanics: A physics of
  and by beliefs. Interface Focus  \textbf{13}(3),  20220029 (2023).
  \doi{10.1098/rsfs.2022.0029}

\bibitem{reynolds1987flocks}
Reynolds, C.W.: Flocks, herds and schools: A distributed behavioral model. In:
  Proceedings of the 14th annual conference on Computer graphics and
  interactive techniques. pp. 25--34 (1987)

\bibitem{talts2018}
Talts, S., Betancourt, M., Simpson, D., Vehtari, A., Gelman, A.: Validating
  bayesian inference algorithms with simulation-based calibration. arXiv
  preprint arXiv:1804.06788  (2018)

\bibitem{waade2025}
Waade, P.T., Olesen, C.L., Laursen, J.E., Nehrer, S.W., Heins, C., Friston, K.,
  Mathys, C.: As one and many: Relating individual and emergent group-level
  generative models in active inference. Entropy  \textbf{27}(2), ~143 (2025).
  \doi{10.3390/e27020143}

\end{thebibliography}

% Supplementary material (appendices) follow on a new page.
\clearpage
% =====================================================================
% Supplementary material for "Collective Sensing as Emergent Bayesian
% Inference". Not a standalone document: it is \input{} into main.tex
% after the bibliography (see the end of main.tex), so all cross-references
% to/from the main text resolve in a single compilation.
% =====================================================================
\appendix

% Preserve appendix names and unique links with newer LaTeX kernels.
\crefalias{section}{appendix}
\renewcommand{\theHsection}{appendix.\arabic{section}}

\section{The joint model fit}\label{sec:results:inverse}

\begin{figure}[t]
\centering
\includegraphics[width=\textwidth]{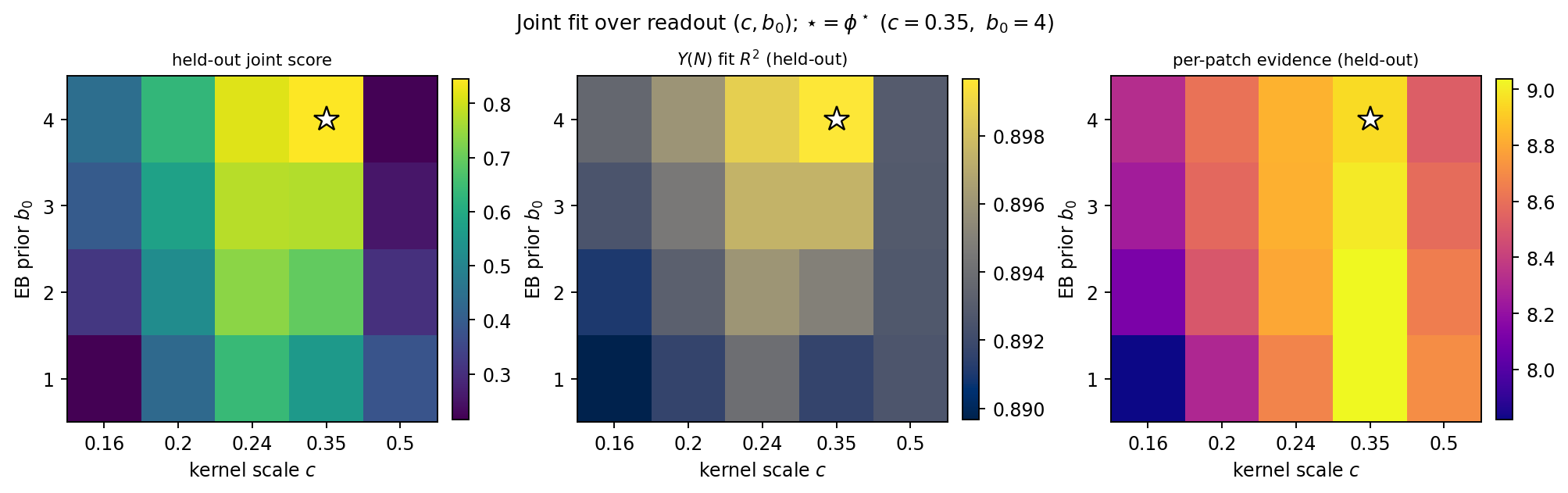}
\caption{The joint fit over the generative model $(c, b_0)$, with the heading
parameters $(\rho,\Sigma_c)$ fitted on training seeds at every point
(star: the validation-selected $\phi^\star$, $c=0.35$, $b_0=4.0$).
The displayed validation scores were also used to select the posterior model;
the patch score is the posterior log density of the true gradient.}
\label{fig:jointfit}
\end{figure}

We fit the generative model to simulated trajectories by comparing
20 posterior-model candidates,
$c\in\{0.16,0.20,0.24,0.35,0.50\}$ and
$b_0\in\{1,2,3,4\}$, fitting heading persistence~$\rho$ and
isotropic coupling variance~$\Sigma_c$ at each point.
The displayed selection is
\[
\phi^\star:\quad c = 0.35,\qquad b_0 = 4.0,
\]
with $\rho=1.25$, $\Sigma_c\approx0.255$, and fixed $\Sigma_h=I$ (\Cref{fig:jointfit}).
Diagnostic variants are listed below.

We use 64 seeds at each group size
$N\in\{4,8,16,32,64,128,256\}$, at $\ell_\mathrm{field}=40$:
448 rollouts with 60 recorded frames after 16 warm-up steps.
Seeds are generated by NumPy's \texttt{SeedSequence(20260421)};
sorting the 64 distinct seed values and assigning odd zero-based ranks
to validation gives 32 training and 32 validation seeds per~$N$.
Each seed retains its split across frames and group sizes. For each
candidate, heading parameters and the affine $Y$ mapping are fitted
on training seeds. Validation scores select the posterior model;
the reported alignment of approximately~$0.59$ and $Y$-fit
$R^2\approx0.90$ therefore include this selection step. The original selection
used grid-step reference gradients in its calibration and posterior log scores;
the calibration diagnostics below are recomputed in arena units with
$\phi^\star$ held fixed.

The joint score combines four quantities, each min--max scaled across
the candidate grid within a split: a calibration-and-cosine score
(weight~$0.20$), the Gaussian posterior log density of the true local
gradient (weight~$0.20$), action alignment above a shuffled-direction
null (weight~$0.30$), and the $Y$-fit $R^2$ (weight~$0.30$).
The first penalizes deviations of the median
error-to-variance ratio from $\ln2$ and low patch cosine; the median
target is exact only for an isotropic Gaussian error model.
Within each group size the posterior scores pool active patch--frame
records; condition-level scores are then averaged equally across~$N$.

The predicted $Y(N)$ curve is constructed by computing the fitted
heading model's mean cosine to the true COM gradient at each~$N$,
then applying a positive affine map fitted to the training condition
means of observed~$Y$. Validation uses the same affine coefficients.
This empirical mapping uses the simulated true gradient.

\section{Simulation and estimation details}\label{sec:supp:simulation}

\Cref{tab:simulation} records the baseline in simulator units. We use the
\texttt{legacy/original} preset with the paper's speed and coupling
overrides, giving equal orientation and attraction weights.

\begin{table}[t]
\centering\small
\caption{Simulation and analysis baseline, before the experiment-specific
overrides stated in the text.}
\label{tab:simulation}
\begin{tabular}{@{}p{0.45\textwidth}p{0.49\textwidth}@{}}
\toprule
Quantity & Value or implementation \\
\midrule
Arena and light grid & $100\times100$; $128\times128$ grid; reflecting school boundaries \\
Social zones $(r_r,r_o,r_a)$ & $(3,20,25)$; neighbors closer than 1 excluded \\
Orientation/attraction weights & $w_\mathrm{ori}=w_\mathrm{att}=0.5$ \\
Repulsion & Priority override; no independent fitted $w_\mathrm{rep}$ \\
Heading noise; maximum turn & Angular Gaussian SD $0.005$ radians; cap $2$ radians/update \\
Wall avoidance; initialization & Wall weight $0.8$; initial disk radius $100/6$ \\
Speed bounds; school step & $v_\mathrm{min}=7.5$, $v_\mathrm{max}=25$; $\Delta t=0.1$ \\
Light response & Speed decreases linearly with frame-normalized darkness \\
Field parameters & $\ell_\mathrm{field}=40$, $k_0=0.5$, $\nu=0.0005$, background amplitude $1$ \\
Radial feature parameters & Amplitude $1.5$, floor $-3$, displacement $0.25$/field update \\
Field time step & $1$; one field update per simulator step \\
Spatial estimator & 4 moving windows; radius factor $0.4$; within-window connectivity radius $r_o/2=10$ \\
Component eligibility & At least 2 agents; maximum centered offset at least $1$ \\
Numerical regularization & Kernel jitter $10^{-6}$; design ridge $0.02$ \\
EB prior and selected model & $a_0=2$; $c=0.35$, $b_0=4$ for headline calibration \\
Temporal update; school prior & Process covariance $10^{-4}I$; weak school precision $10^{-3}I$ \\
\bottomrule
\end{tabular}
\end{table}

The stored \texttt{light\_field} encodes darkness $D$ in \eqref{eq:light}.
With $f_\mathrm{max}=1.5$ and $f_\mathrm{min}=-3$, the floor is inactive and the positive feature is dark.
Speed is $v_\mathrm{max}-(v_\mathrm{max}-v_\mathrm{min})(D-D_\mathrm{min})/(D_\mathrm{max}-D_\mathrm{min})$, using current-frame extrema.
Scoring instead normalizes darkness over the trajectory and reference fields jointly;
$L_\mathrm{norm}$ denotes its complement. The recorded production code averages
full independently seeded fields for the reference, without explicitly removing the radial feature.

The main calibration curve uses 16 seeds per~$N$ at decay length~40;
the behavior, field-mismatch, and joint-fit analyses use 64 seeds per~$N$.
The structured-field illustration instead uses $N=96$, seed~20260607,
frame~33, $\ell_\mathrm{field}=48$, $k_0=2\pi/22$, background amplitude
$60$, radial amplitude~$1$, and $b_0=0.5$.
The field-noise sweep uses $N=64$, 16 seeds per cell,
$\ell_\mathrm{field}=30$, feature displacement~$0.5$,
$k_0\in\{0.25,0.35,0.5,0.75,1\}$,
$\nu\in\{0.0002,0.0003,0.0005,0.00075,0.001\}$, and the
recorded regeneration configuration has $b_0=0.5$.
The patch-local shuffle and overlap controls use $\phi^\star$,
$N=64$, 16 seeds and $\ell_\mathrm{field}=40$; the earlier graded
controls use the iid likelihood (Supplementary \Cref{sec:supp:doseresponse}).

The social-off control sets orientation and attraction weights to zero,
preserving speed modulation, walls, heading noise, and the short-range
repulsion rule. When neighbors occupy only the
orientation or attraction zone, that directional branch remains active.
The alignment-fraction sweep changes $(w_\mathrm{ori},w_\mathrm{att})$
to $(r,1-r)$, with $r\in\{0,0.25,0.5,0.75,1\}$.

\paragraph{Likelihood assumptions and the simulator--model connection.}
\label{sec:supp:assumptions}
The likelihood requires locally linear field and speed responses and
approximately Gaussian residuals with the stated covariance. Weighted
geometry determines precision; identifying both gradient components
requires non-collinear positions, while the ridge stabilizes unmeasured directions.

Moving windows select agents; speeds and positions are centered within each
window's connected components. The RBF covariance is block-diagonal across
components, with length set by component RMS spread. Overlapping windows
and consecutive frames reuse agents and correlated field values, limiting
conditional independence. Pooling also assumes coherent local gradients;
components on opposite sides of a curved feature can violate this assumption.

Alignment, limited turning, and isolated-heading retention motivate persistence.
The physical path is gradient $\to$ speed differences $\to$ relative
positions and social turning $\to$ heading. We hypothesize that these
collective dynamics realize approximate inference and belief-guided action
under the model's assumptions. The generative model specifies statistical
relations for inference; its sampling procedure and graphical arrows may
differ from the processes generating the school's physical trajectories.

One operational test fixes a physical coarse-graining $C$ of school states
$s$, inputs $y$, and an approximate-inference update $B_y$. Physical evolution
$F_y$ should satisfy $C(F_y(s))\approx B_y(C(s))$ within a stated regime and
error tolerance (in distribution for stochastic updates). Here $C$ reads
collective belief parameters. Fixed $C$ and $B_y$ permit derivations or
perturbation tests across states and inputs. Establishing this correspondence
requires identifying a collective state for the replay filter's memory.
Conditional Gaussian heading updates are exact under the assumed
approximation; gradient estimation uses plug-in EB precision (\Cref{app:iid}).

The implemented replay also filters each spatial window through time.
It initializes patch precision at $4I$, predicts
$\Sigma^-_{p,t}=\Sigma_{p,t-1}+10^{-4}I$, and combines this prior with the current Gaussian likelihood. School fusion adds complete active filtered patch precisions and information vectors, plus $10^{-3}I$ school precision, retaining duplicated patch priors and shared past evidence. The static overlap control (\Cref{sec:supp:doseresponse}) applies Equation~\eqref{eq:federation} to current likelihood messages.

\section{Expected free energy and the posterior-gradient drive}\label{sec:supp:efe}

The action drive $u_t = W_t\mu_\mathrm{school}$ of~\eqref{eq:action}
enters the conditional Gaussian heading update~\eqref{eq:heading-update}.
We derive its relationship to a one-step expected-free-energy (EFE) objective.

For a local, one-step, quasi-static approximation, treat the unit
heading $a$ as the school's action, with preferences over the
predicted darkness. Moving a unit step along $a$ changes the local light level by
$-\mu_\mathrm{school}^\top a$ to first order, so the predicted next
observation is Gaussian,
\[
q(o\mid a) = \mathcal{N}\!\big(L_0 - \mu_\mathrm{school}^\top a,\; V(a)\big),
\qquad V(a) = a^\top\Sigma_\mathrm{school}\,a + s^2,
\]
with $L_0$ the current mean light level at the school and $s^2$ irreducible
observation noise; the variance inherits the posterior uncertainty
$a^\top\Sigma_\mathrm{school}\,a$. The
school prefers a dark target $L^\star$ with tolerance $\sigma_C^2$, $p(o\mid
C) = \mathcal{N}(L^\star,\sigma_C^2)$. For this homoscedastic likelihood the ambiguity term of
the EFE is action-independent, so minimizing the EFE reduces to minimizing
the risk,
\[
R(a) = D_\mathrm{KL}\!\big[\,q(o\mid a)\,\big\|\,p(o\mid C)\,\big]
= \tfrac12\Big[\!-\!\ln V(a) + \tfrac{V(a)}{\sigma_C^2}
+ \tfrac{(D-\mu_\mathrm{school}^\top a)^2}{\sigma_C^2}\Big] + \mathrm{const},
\]
where $D = L_0 - L^\star$ is the darkness deficit. Stationarity on the unit
sphere, with Lagrange multiplier $\lambda$, requires
\[
(\alpha\,\Sigma_\mathrm{school} + \lambda I)\,a^\star = \beta\,\mu_\mathrm{school},
\qquad
\alpha = \tfrac{1}{\sigma_C^2} - \tfrac{1}{V^\star},\quad
\beta = \tfrac{D - \mu_\mathrm{school}^\top a^\star}{\sigma_C^2},
\]
where $V^\star=V(a^\star)$ and $\lambda$ enforces unit length.
When $\alpha>0$, $\beta>0$, and $\lambda/\alpha>0$, this condition
has the regularized precision-weighted form
\[
a^\star\;\propto\;
\big(\Sigma_\mathrm{school}+(\lambda/\alpha)I\big)^{-1}
\mu_\mathrm{school}.
\]
Identifying $\lambda/\alpha$ with the fitted coupling variance $\Sigma_c$
gives the algebraic form of $u_t$. This approximation relates a fitted
parameter to a multiplier determined by preferences, observations and
the unit-length constraint. The conditions $V^\star>\sigma_C^2$ and
$\lambda/\alpha>0$ require separate checks, as does global minimality.

This construction neglects field change during a step, curvature
across the displacement, future information gathering, and the social
and turning constraints on feasible motion. With vanishing posterior
mean, motion can persist through the heading prior. The empirical test
measures displacement alignment with the full predicted heading
(\Cref{sec:results:efe}); testing EFE optimization by the collective
dynamics remains an extension.

\section{Calibration}\label{sec:supp:calibration}

\Cref{sec:results:calibration} adopts the endogenous school-coupled
kernel, motivated by the overconfidence of simpler likelihoods in
simulator data. The iid centered-residual variant (\Cref{app:iid})
removes the duplicated pairwise evidence but still treats per-agent
residuals as independent. In the simulator these
residuals are not independent: the local light field is spatially
correlated, while attraction and alignment shape the positions at which
agents sample it. The component-dependent length $\ell_{p,k}$
of~\eqref{eq:rbf-kernel_and_ell-eff} accounts for part of this dependence.

\begin{figure}[t]
\centering
\begin{subfigure}{0.62\textwidth}
\centering
\includegraphics[width=\linewidth]{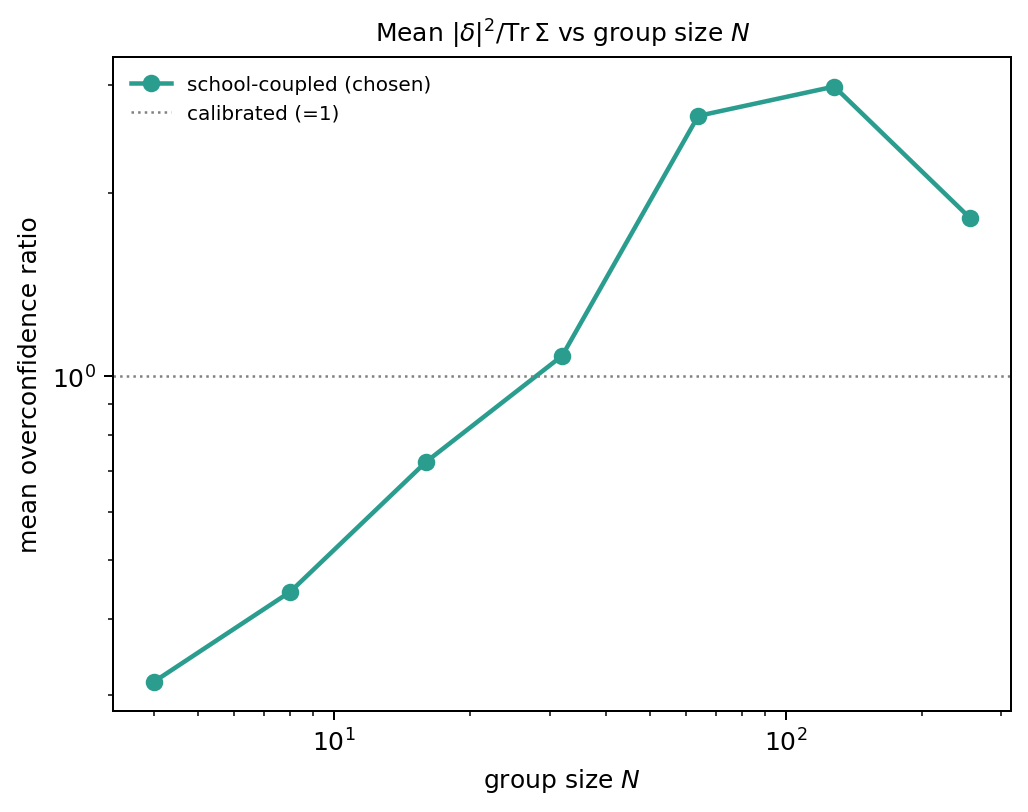}
\caption{}
\label{fig:calibration:overconf}
\end{subfigure}\hfill
\begin{subfigure}{0.36\textwidth}
\centering
\includegraphics[width=\linewidth]{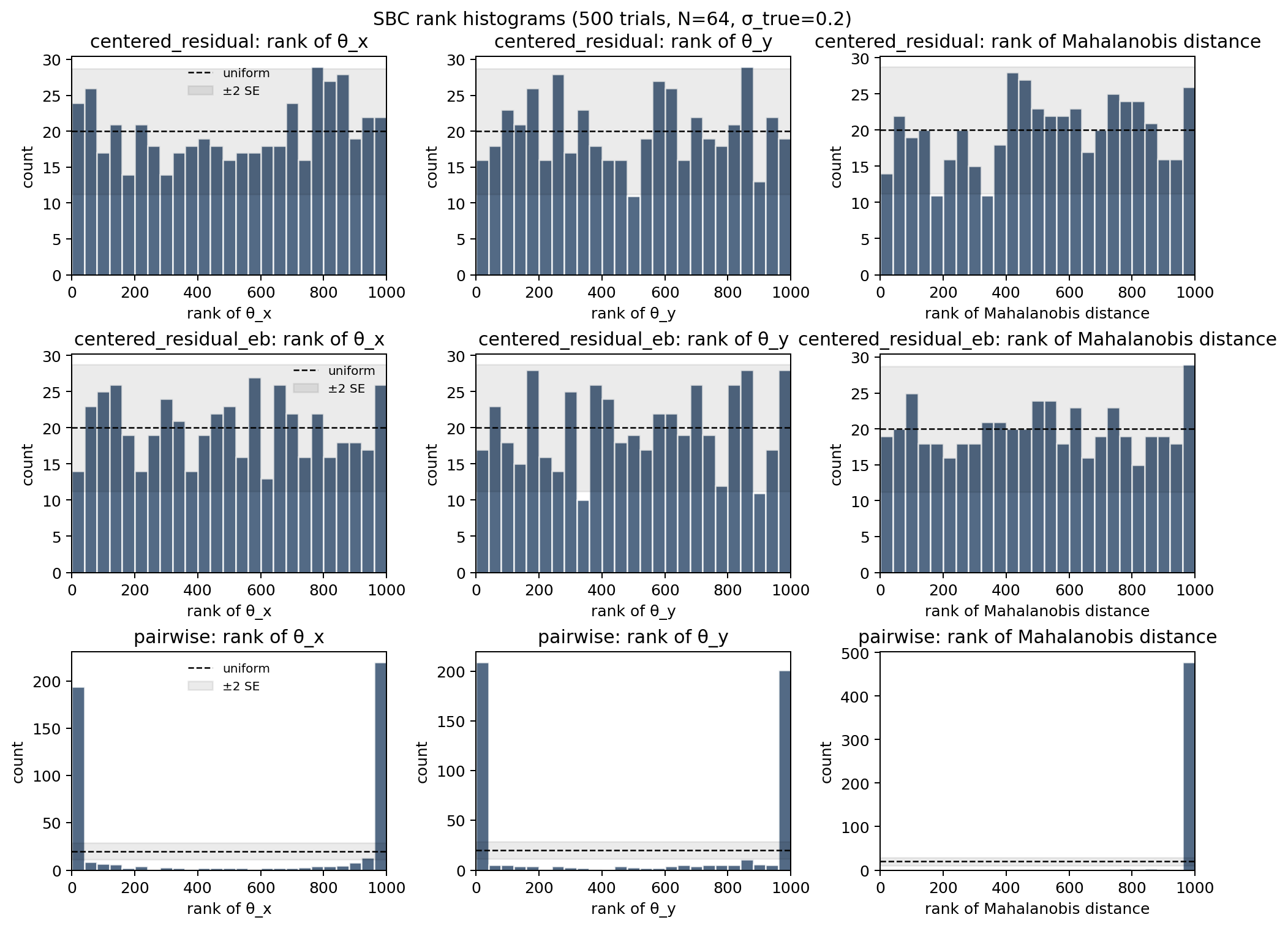}
\caption{}
\label{fig:calibration:sbc}
\end{subfigure}
\caption{Calibration diagnostics.
\textbf{(\subref{fig:calibration:overconf})}~Mean posterior
patch error-to-variance ratio vs.\ group size for the selected
school-coupled posterior; dotted line: mean calibration target~$1$.
\textbf{(\subref{fig:calibration:sbc})}~Synthetic rank histograms,
with a uniform-rank reference, for the simpler estimators at fixed $N=64$.}
\label{fig:calibration}
\end{figure}

The patch-level calibration diagnostic is
$r_p=\|\theta_{p,\mathrm{true}}-\mu_p\|^2/\mathrm{Tr}\,\Sigma_p$,
using the simulator gradient sampled at the patch center. The main
curve and \Cref{fig:calibration}a average ratios over finite active
patch--frame records pooled across 16 rollouts per group size.
Consecutive frames and overlapping windows are dependent. The boundary map first
averages within each rollout and then across seeds. Posterior cosine is
$\mu_p^\top\theta_{p,\mathrm{true}}/
(\|\mu_p\|\,\|\theta_{p,\mathrm{true}}\|)$; degenerate vectors are
excluded. A ratio below one indicates excess posterior variance
relative to error; cosine measures directional accuracy.

Reference derivatives and regression positions use arena coordinates throughout
these diagnostics. Finite differences include the grid spacing
$\Delta x=\Delta y=100/127$ on the $128\times128$ grid. We recompute
estimation errors from the corrected reference vectors before averaging the
ratios; the posterior means and covariances are unchanged.

The synthetic rank check in \Cref{fig:calibration}b follows the rank-based
approach of simulation-based calibration~\cite{talts2018}, using
500 independent synthetic regressions with $N=64$,
positions uniform on $[-5,5]^2$, gradient draws
$\mathcal N(0,0.5^2I)$, speed-noise SD~$0.2$, and 999 Gaussian
posterior draws per trial. The estimator prior is disabled and the
noise variance is fixed in the synthetic generator. Coordinate and
Mahalanobis-distance ranks compare the centered-residual,
centered-residual EB, and pairwise implementations under simple synthetic
noise. A full prior-predictive SBC check of the selected school-coupled
model remains an extension.

\section{Sensitivity to spatiotemporal field noise}\label{sec:results:fieldnoise}

\begin{figure}[t]
\centering
\includegraphics[width=0.52\textwidth]{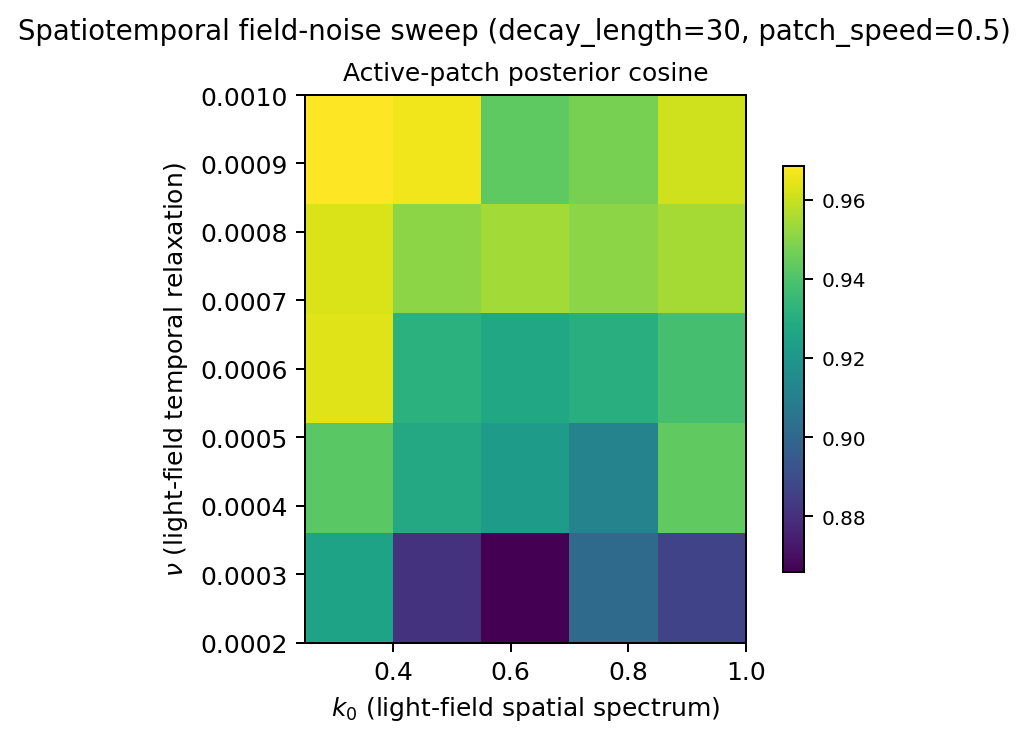}
\caption{Spatiotemporal field-noise sweep: active-patch posterior
cosine as a heatmap over~$(k_0, \nu)$ at fixed
$\ell_\mathrm{field} = 30$ and radial-feature displacement $0.5$ per field update, $N = 64$, 16 seeds per
cell.}
\label{fig:fieldnoise}
\end{figure}

\Cref{fig:fieldnoise} maps a~$5 \times 5$ grid of $(k_0, \nu)$ --
light-field spatial spectrum and temporal relaxation -- onto active-patch posterior cosine.
Posterior cosine remains above~$0.9$ across the reported sweep.

\section{Calibration is environment-relative}\label{sec:results:boundary}

These maps and the subsequent boundary probes use arena-unit reference
gradients, with the selected likelihood parameters held fixed
(\Cref{sec:supp:calibration}). The boundary sweep uses the original settings
and seeds, reusing available cached trajectories and regenerating the remaining
conditions with JAX float32 on CPU.

\begin{figure}[t]
\centering
\includegraphics[width=0.8\textwidth]{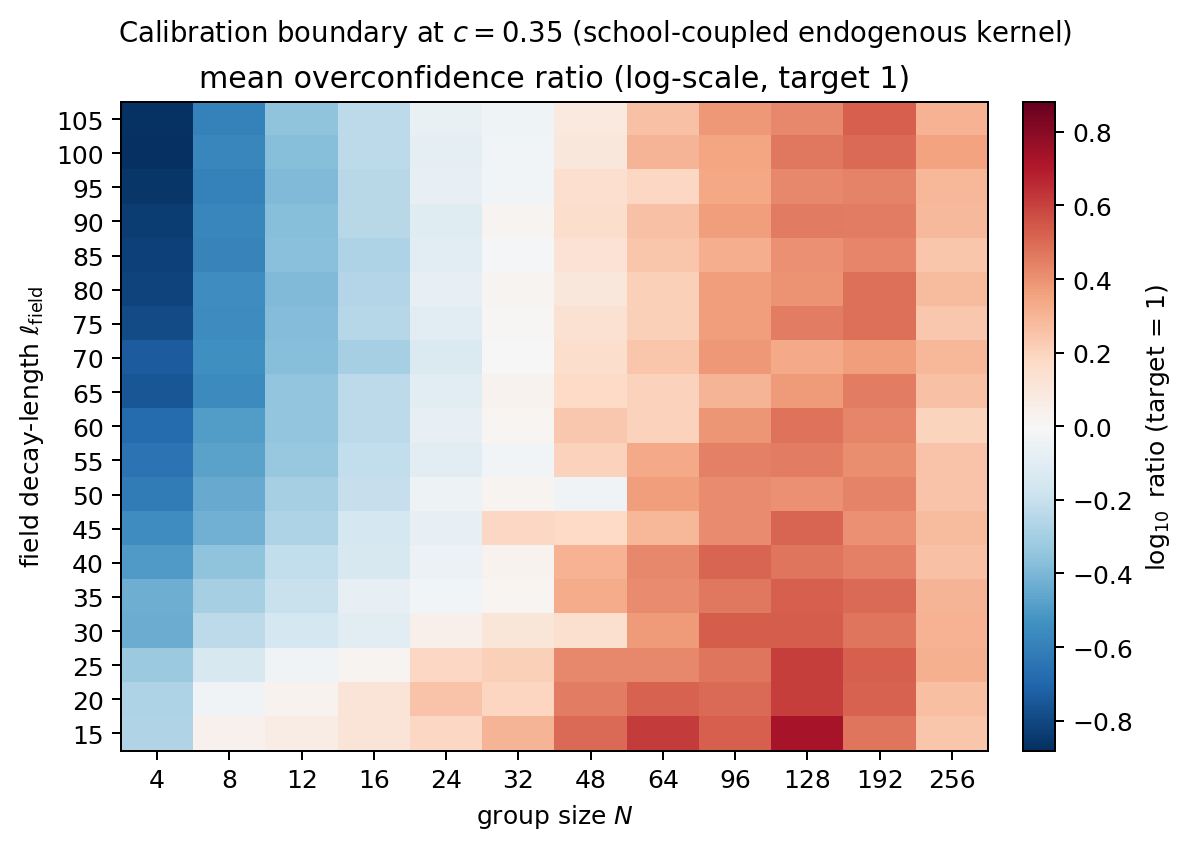}
\caption{Calibration boundary in $(N, \ell_\mathrm{field})$ space
under the school-coupled endogenous kernel (\Cref{eq:rbf-kernel_and_ell-eff})
with $c\!=\!0.35$ and $b_0\!=\!4.0$, 16 seeds per cell across the
12-$N$ $\times$ 19-$\ell_\mathrm{field}$ grid.
Cell color: $\log_{10}$ of the mean patch error-to-variance ratio
relative to the mean calibration target~$1$
(red = over-confident, blue = under-confident). Calibration depends jointly
on school size and environment.}
\label{fig:boundary}
\end{figure}

\Cref{fig:boundary} maps the mean overconfidence ratio at
$c\!=\!0.35$, $b_0\!=\!4.0$, 16 seeds per cell, across
\begin{align*}
N \in &\;\{4,8,12,16,24,32,48,64,96,128,192,256\}, \\
\ell_\mathrm{field} \in &\;\{15,20,25,\ldots,105\}
\end{align*}
For a correctly centered Gaussian error model with the stated
covariance, the mean error-to-variance ratio is~$1$ even when the
covariance is anisotropic. Only in the isotropic two-dimensional case
does the full ratio distribution reduce to
\[
\frac{|\theta_{p,\mathrm{true}} - \mu_p|^2}
     {\mathrm{Tr}\,\Sigma_p}
\;\sim\; \mathrm{Exp}(1) \;\equiv\; \chi^2_2 / 2,
\]
with mean $1$ and median $\ln2\approx0.69$. For $\ell_\mathrm{field}\geq40$, the first sampled size with a mean ratio above one is $N=32$--$48$. At decay lengths $15,20,25,30,35$, this occurs at $N=8,12,16,24,32$, respectively. Ratios peak at $2.4$--$5.3$ across decay lengths, at $N=96$--$192$, and remain above one ($1.6$--$2.3$) at $N=256$. Spatial decay jointly changes local curvature and residual correlation. The wider EB prior ($b_0=4$ versus the class default $0.5$) allows greater residual variance; $c$ remains fixed across environments.

\section{Effects of contrast and spatial scale}\label{sec:results:falsification}

We tested two environmental regimes with fixed agent rules and the
school-coupled EB model of \Cref{fig:boundary}.
In C1, we scaled both amplitude and floor parameters of
the radial light feature by factors $1,2,4,8,16$. This increases its
amplitude relative to the background while keeping its decay length
fixed. In C2,
we pushed large schools beyond the upper edge of
the calibration boundary while varying the field decay length.

Field curvature within components, inconsistent gradients across
components, and error in the positional covariance kernel can each
produce miscalibration. Decreasing $\ell_\mathrm{field}$ can affect all three;
increasing contrast changes signal, curvature magnitude, and the
normalization of speed sensitivity together. C1 and C2 therefore probe
model adequacy without separately identifying these mechanisms.

\begin{figure}[t]
\centering
\includegraphics[width=\textwidth]{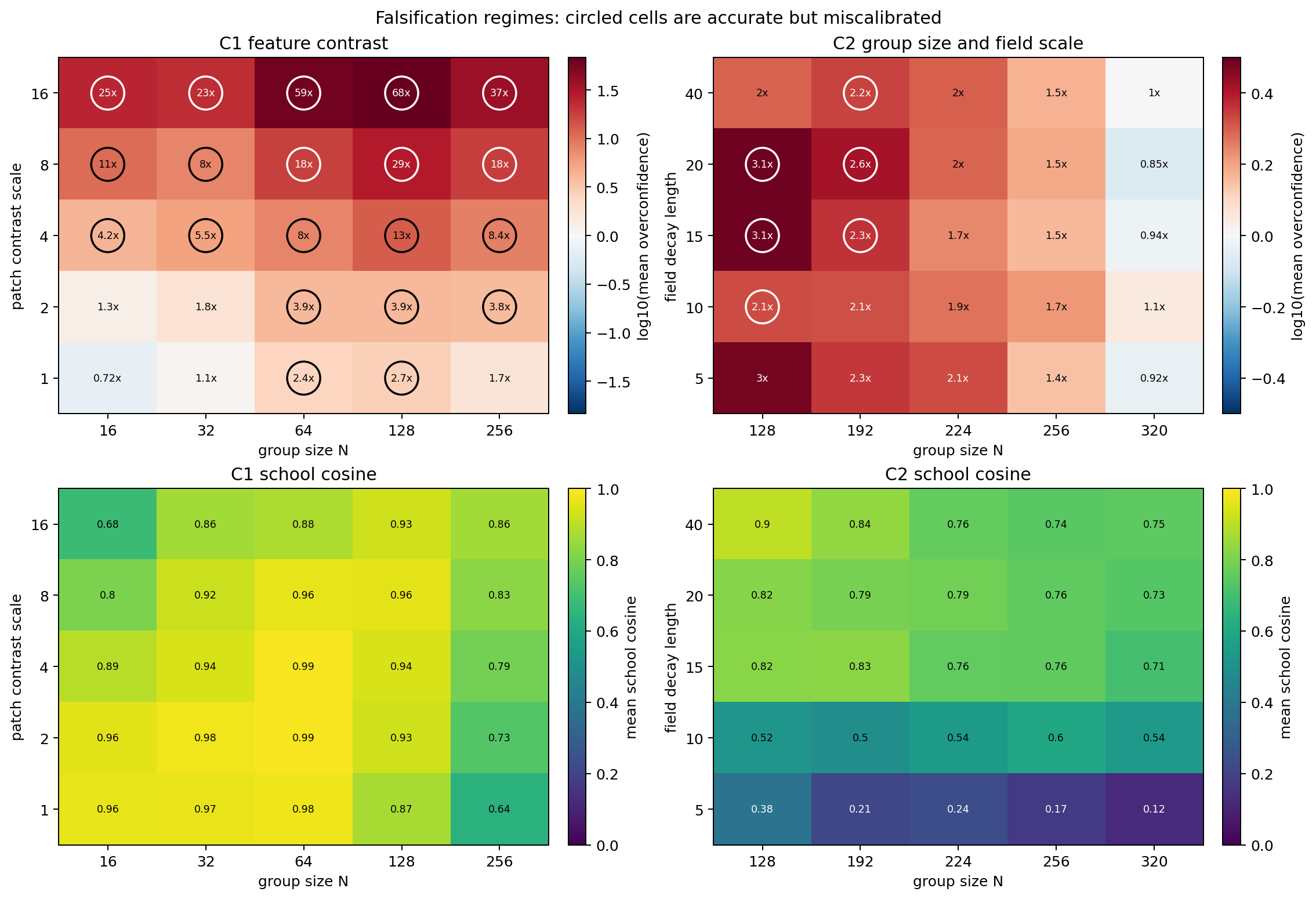}
\caption{Boundary probes for the school-coupled EB posterior.
Top row: mean patch error-to-variance ratio relative to its calibration target of $1$; red cells are over-confident
and blue cells are under-confident. Bottom row: school-level posterior
cosine to the true local gradient. Circled cells have mean school
cosine $>0.5$ and mean overconfidence outside a factor-of-two
calibration band.
\textbf{(a)}~C1 contrast scaling: high
contrast preserves school-level directional accuracy but makes patch posteriors too confident. \textbf{(b)}~C2: calibration varies with school size and spatial scale;
the shortest decay length also reduces directional accuracy.}
\label{fig:falsification}
\end{figure}

\Cref{fig:falsification} shows how the posterior changes across these
regimes. At baseline contrast, the C1 mean error-to-variance ratio ranges
from $0.72$ to $2.72$, with $N=64$ and $N=128$ above the factor-of-two
band around the nominal target. Raising contrast to
$4\times$--$16\times$ preserves positive direction estimates (school
cosine $0.68$--$0.99$) while increasing the mean ratio to $4.2$--$68$. In C2, schools of $N=128$--$192$ at
$\ell_\mathrm{field}=15$--$20$ remain directionally accurate (school
cosine $0.79$--$0.83$), with error-to-variance ratios of $2.30$--$3.07$.
At $N=320$ and $\ell_\mathrm{field}=15$--$40$, the mean ratio is
$0.85$--$1.00$, close to the nominal target, while school cosine is
$0.71$--$0.75$. The shortest decay length ($5$) reduces school cosine
to $0.12$--$0.38$ across the tested sizes.

\section{Field-specificity of the group-size law}\label{sec:supp:fieldmismatch}

We test whether the rise of~$Y$ with~$N$ (\Cref{sec:results:berdahl})
depends on the particular field the school experienced. Holding cached
trajectories fixed, we recompute~$Y$ against a field from an independent
seed (\emph{cross-seed}) and compare it with the true field
(\emph{natural}) and the $Y=1$ reference.

\begin{figure}[t]
\centering
\includegraphics[width=0.6\textwidth]{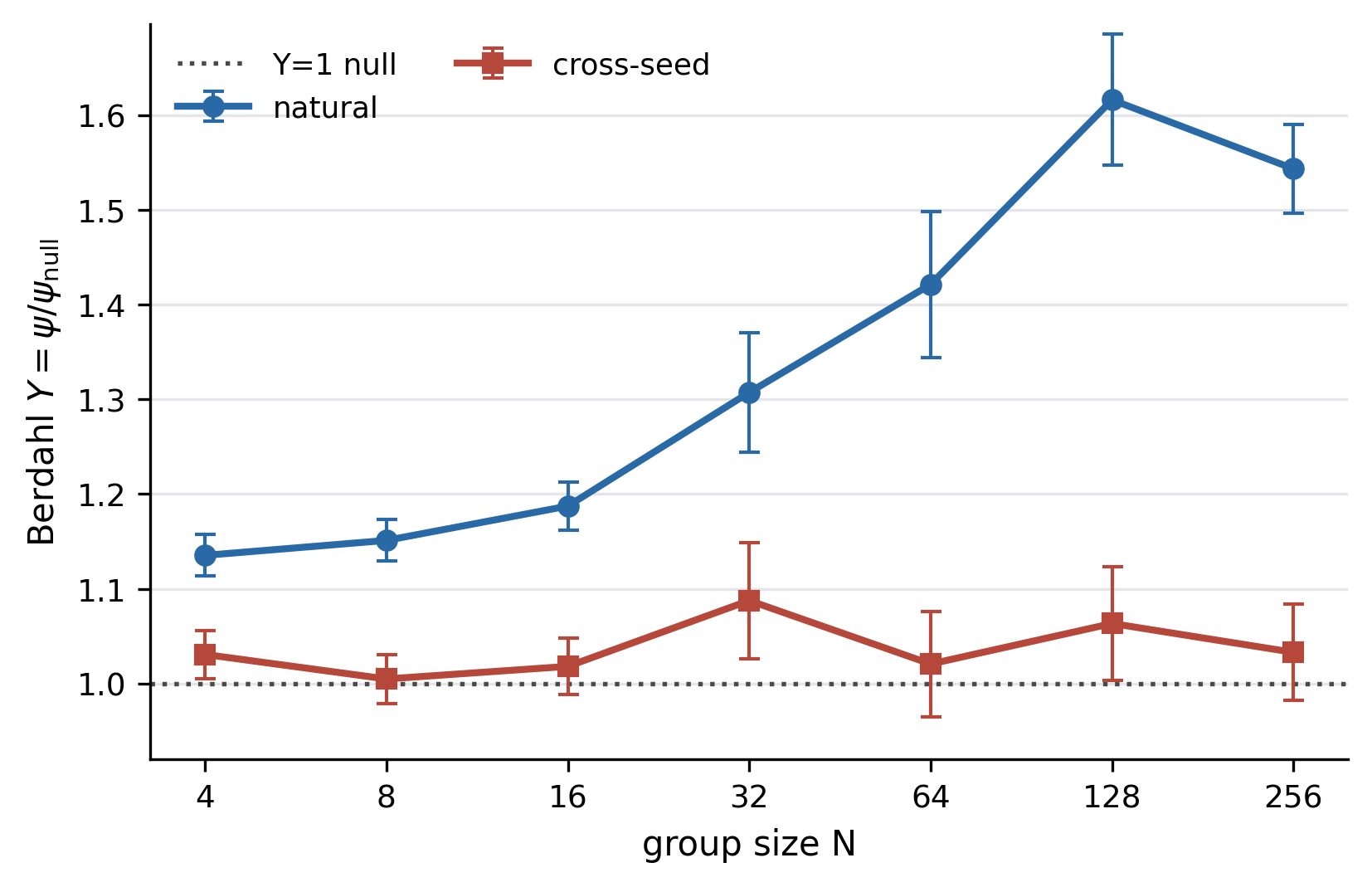}
\caption{The group-size performance law is field-specific. $Y$
vs.~$N$ for the cached school trajectories scored against the true field
(\emph{natural}) and an independent seed's field (\emph{cross-seed}),
with the $Y=1$ reference. The positive $Y$--$N$ slope survives only
for the true field and vanishes under field decoupling.}
\label{fig:fieldmismatch}
\end{figure}

Against the true field, $\overline{Y}\approx1.3$ with slope $+0.085$
per doubling of~$N$ (\Cref{fig:fieldmismatch}). Against an independent
field, $\overline{Y}\approx1.0$ and the slope falls to $+0.004$.
The school therefore preferentially occupies dark regions of the field
that generated its motion~\cite{berdahl2013}. This environmental coupling
is a requirement for the Bayesian-mechanics
interpretation~\cite{dacosta2021,ramstead2023}; the inference model is
assessed separately through its gradient and heading predictions.

\section{Alignment-fraction dependence of $Y$}\label{sec:supp:berdahl}

\begin{figure}[t]
\centering
\includegraphics[width=0.6\textwidth]{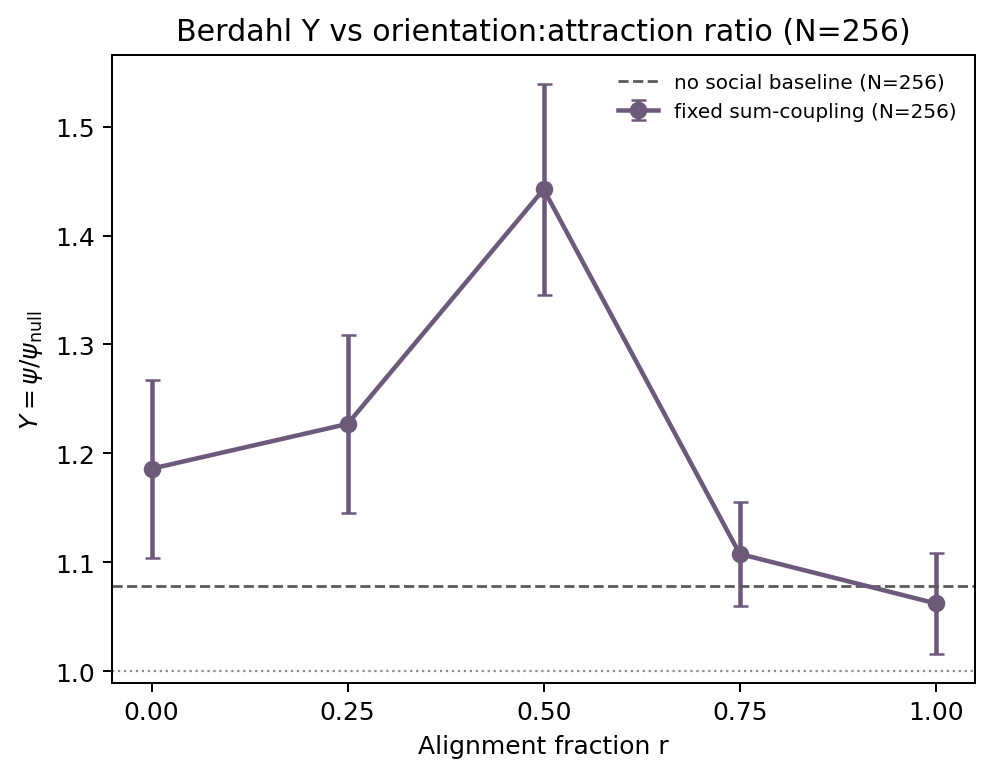}
\caption{Darkness-exposure ratio~$Y$ vs.\ the alignment fraction
$r = w_\mathrm{ori}/(w_\mathrm{ori}+w_\mathrm{att})$ at fixed
sum-coupling and the largest group size ($N=256$), with the social-off
baseline (dashed) and the $Y=1$ null (dotted). $Y$ peaks near $r=0.5$,
where alignment and attraction balance.}
\label{fig:berdahl-r}
\end{figure}

The ratio~$Y$ also varies with the alignment fraction
$r = w_\mathrm{ori}/(w_\mathrm{ori}+w_\mathrm{att})$ at fixed
sum-coupling (\Cref{fig:berdahl-r}): $Y$ peaks at intermediate~$r$,
where alignment and attraction balance, reproducing the known
Couzin-rule trade-off~\cite{couzin2002,berdahl2013}. Too little
alignment and the school fails to polarise; too much and it loses the
cohesion that supports coordinated movement through the field.

\section{Patch-local evidence and overlap controls}\label{sec:supp:doseresponse}

We test the contribution of local speed--position relationships by shuffling speeds within connected components or whole spatial windows before centering. Each intervention preserves the speed distribution at its chosen scale; whole-window shuffling also exchanges speeds between disconnected components. We apply the selected model $\phi^\star$ ($c=0.35$, $b_0=4$) to 16 cached $N=64$ trajectories at $\ell_\mathrm{field}=40$, retaining temporal filtering and 60 recorded frames after 16 warmup frames. Trajectories are fixed, and permutations are drawn independently for each frame and window. An agent in overlapping windows can therefore receive different shuffled speeds in each local inference calculation. These trajectories previously contributed to the calibration diagnostics. Reference gradients in both controls use arena-unit derivatives.

\begin{figure}[tbp]
\centering
\includegraphics[width=\textwidth]{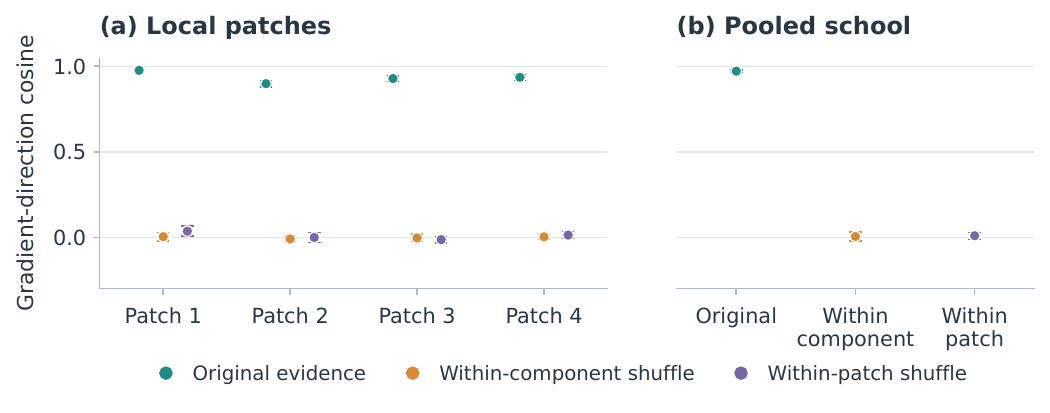}
\caption{Local evidence controls at the selected correlated model ($N=64$). Cosine to the true gradient at \textbf{(a)} each window center and \textbf{(b)} the school COM. Speeds are shuffled within connected components or whole windows. Points show means over 16 seed summaries; bars show 95\% seed-bootstrap intervals. Each shuffled seed summary averages eight repetitions.}
\label{fig:doseresp}
\end{figure}

Mean active-patch cosine falls from $0.935$ to $-0.001$ under component shuffling and $0.009$ under whole-window shuffling; school cosine falls from $0.971$ to $0.006$ and $0.010$, respectively (\Cref{fig:doseresp}). We average frames and eight shuffle repetitions within each seed, then use 10,000 seed-bootstrap resamples for percentile 95\% intervals. The paired whole-window-minus-component difference is $0.010$ ($[-0.005,0.026]$) for patch cosine and $0.004$ ($[-0.033,0.038]$) for school cosine. Both interventions remove almost all directional accuracy, with a small positive residual in patch~1 under whole-window shuffling ($0.037$, $[0.009,0.066]$). All four windows are active on all 960 frames, and all cosine values are defined.

\paragraph{Overlap sensitivity.} We fix the set of covered agents and assign each agent to its nearest eligible window center (ties by window index). Coverage averages $54.9$ of 64 agents; the original windows duplicate $41.0\%$ of covered agents, with mean membership multiplicity $1.43$. All four windows remain active under unique allocation. Mean agent count per window falls from $19.6$ to $13.7$. We compare static likelihood messages with the weak school prior added once, excluding temporal filtering. Targets are unchanged window centers and school COM. Nominal ellipse coverage tests $(\theta-\mu)^\top\Lambda(\theta-\mu)\leq5.991$.

Unique allocation increases patch coverage from $87.0\%$ to $92.1\%$ (paired improvement $5.1$ percentage points, 95\% interval $[3.4,6.8]$), while the mean error-to-variance ratio falls from $1.84$ to $0.91$ and patch cosine rises from $0.943$ to $0.949$. School coverage increases from $46.8\%$ to $53.3\%$, with ratios $15.97$ and $10.39$ and cosines $0.963$ and $0.972$. Common-active and native pooling coincide here. Larger posterior uncertainty contributes to the school coverage gain; the paired interval for the change in school squared error spans zero. Coverage varies between patches, and substantial school-level overconfidence remains. Repartitioning changes geometry, component sizes and kernel scales alongside overlap. Gradients also vary across the school: mean true patch-to-COM cosine is $0.714$, with relative gradient-vector difference $0.577$. These features affect the conditional-independence and shared-gradient assumptions underlying pooling.

\paragraph{Earlier graded controls.} The iid controls use $N=64$, 16 seeds, decay length~30 and $\alpha\in\{0.1,0.25,0.5,0.75,0.9\}$. Attenuation uses $\bar v+\alpha(v_i-\bar v)$, where $\alpha=1$ is intact evidence. The code's \texttt{uniform\_noise} adds Gaussian noise with component variance $(1-\alpha)^2 2s_v^2m/(m-1)$ for $m>1$ agents. At $\alpha=0$, this matches full shuffle's expected post-centering perturbation variance.

\section{The iid centered-residual posterior}\label{app:iid}

For the $K=I$ reduction of \eqref{eq:school-coupled-likelihood}, let
$Z$ include the signed speed sensitivity times the centered positions,
$y$ collect centered speeds, and $A=Z^\top Z+\tau I$, with numerical
ridge $\tau=0.02$. The implemented likelihood message is
$\mu=A^{-1}Z^\top y$, $\Lambda=(a_M/b_M)A$, where
$a_M=a_0+M/2$ and
$b_M=b_0+\tfrac12\max\{y^\top y-\mu^\top A\mu,0\}$;
$M$ counts retained agent residuals across components in the window.
The Gaussian message uses mean noise precision $a_M/b_M$, rather than
the variance-marginalized Student-$t$ posterior. Centering constrains the
residuals; $M$ is a row count, not an effective sample size.

\section{Centering and the GLS framing of the posterior}\label{app:centering}

The component nuisance mean is removed by arithmetic centering before
the RBF weights are applied. If raw residual covariance were $C$,
the exact centered covariance would be $HCH^\top$,
$H=I-\mathbf1\mathbf1^\top/m$, which is singular on the constant
direction. The implementation instead uses a positive-definite RBF
matrix with diagonal jitter as a \emph{working} covariance for the
centered residuals. Consequently it is not an exact marginalization
of a raw-speed Gaussian model with unknown intercept.
Full GLS would estimate the nuisance mean using
$\hat\mu_c=(\mathbf1^\top C^{-1}\mathbf1)^{-1}
\mathbf1^\top C^{-1}v$ and residualize the design consistently.
Arithmetic and covariance-weighted centering agree when $\mathbf1$
is a covariance eigenvector; the endogenous length rule alone does
not guarantee this condition.

\end{document}